\documentclass[11pt,a4paper]{article}
\usepackage{amssymb,amsmath,palatino,amsfonts}
\usepackage{graphicx,subfigure}

\newcommand\be{\begin{equation}}
\newcommand\ee{\end{equation}}
\newcommand\bea{\begin{eqnarray}}
\newcommand\eea{\end{eqnarray}}
\newcommand\ket[1]{|#1\rangle}
\newcommand\bra[1]{\langle #1|}

\begin{document}

\title{{\large \textbf{Constructing Mutually Unbiased Bases in Dimension Six}}}
\author{{\normalsize Stephen Brierley and Stefan Weigert} \\
{\normalsize Department of Mathematics, University of York}\\
{\normalsize Heslington, UK-York YO10 5DD}\\
{\normalsize \texttt{sb572@york.ac.uk,slow500@york.ac.uk}}}
\date{{\normalsize April 2009}}
\maketitle

\begin{abstract}

The density matrix of a qudit may be reconstructed with \emph{optimal} efficiency if the expectation values of a specific set of observables are known. In dimension six, the required observables only exist if it is possible to identify \emph{six} mutually unbiased complex ($6\times 6$) Hadamard matrices. Prescribing a first Hadamard matrix, we construct all others mutually unbiased to it, using algebraic computations performed by a computer program. We repeat this calculation many times, sampling all known complex Hadamard matrices, and we never find more than \emph{two} that are mutually unbiased. This result adds considerable support to the conjecture that no seven mutually unbiased bases exist in dimension six.
\end{abstract}

\section{Introduction}

Suppose you want to reconstruct the density matrix $\rho$ of a qudit, a
quantum system with $d$ orthogonal states. To apply the \emph{most efficient} reconstruction method you should measure a set of observables associated with $d$ mutually unbiased complex Hadamard matrices of size $(d\times d)$. A \emph{complex Hadamard matrix} $H$ is a unitary matrix having entries of modulus $1/\sqrt{d}$ only; two such matrices are said to be \emph{mutually unbiased} (MU) if their product is another Hadamard matrix, 
\begin{equation}  \label{MUH}
H^{\dagger }H^{\prime }=H^{\prime \prime } \, , 
\end{equation}
where $H^{\dagger }$ denotes the adjoint of the matrix $H$. The columns of 
$d $ MU complex Ha\-da\-mard matrices and of the identity define a \emph{complete} set
of $(d+1)$ MU orthonormal bases in the space $\mathbb{C}^{d}$ suitable for optimal state reconstruction  \cite{ivanovic81,wootters+89}. Such bases are often characterized directly by the scalar products between their elements,
\begin{equation} \label{MUB conditions}
\left\vert {\langle \psi _{j}^{b}|}{\psi _{j^{\prime }}^{b^{\prime }} 
            \rangle }\right\vert 
            =\left\{ 
            \begin{array}{ll} \delta_{jj^{\prime }} & \quad 
            \mbox{if $b = b^\prime$}\, , \\ 
            \frac{1}{\sqrt{d}} & \quad 
            \mbox{if $b \neq b^\prime $}\,,\end{array} \right.  
\end{equation}
where $b,b^{\prime }=0,1,\ldots ,d$. MU bases they are also useful for quantum cryptography \cite{cerf+02} and play an important role in the solution of the Mean King's problem \cite{Aharonov+01}.

Here is the catch: as of today, complete sets of MU bases have been
constructed only in spaces $\mathbb{C}^d$ of prime or prime power
dimension. If the dimension is a \emph{composite} number, $d=6,10,12, \ldots$, 
the existence of a complete set of MU bases in $\mathbb{C}^d$ has neither
been proved nor disproved (see \cite{planat+06} for a review). In other words, it is unknown even for a qubit-qutrit system whether there exists a set of observables which would realize optimal state reconstruction. Interestingly, constructing a complete set of MU bases in $\mathbb{C}^d$ is equivalent to finding an orthogonal decomposition of the Lie algebra $sl_d(\mathbb{C})$. This poses a long-standing open problem whenever $d$ is not a prime power \cite{boykin+07}.

Let us summarize what is know about the (non-) existence of MU bases in composite dimensions. There are a few \emph{analytic} results: 

\begin{itemize}
\item it is possible to construct \emph{three} MU bases in $\mathbb{C}^d$ without  reference to the value of $d$ \cite{grassl04}; hence, three MU bases do exist for any composite dimension $d$;
\item there are at least $(p^k+1)$ MU bases where the $p^k$ is the smallest factor in the prime decomposition of $d$ \cite{klappenecker+04};
\item more than $(p^k+1)$ MU bases are known exist for specific values of $d$; for example, if $d=2^2 \times 13^2$, a total of $6$ $(\equiv p^k + 2)$ MU bases have been identified \cite{wocjan+05}.
\end{itemize}
Attempts to generalize number-theoretic formul\ae\ used in the construction of complete MU bases from prime-power dimensions  to composite dimensions fail \cite{archer05}. Furthermore, searches for MU bases in dimension six by \emph{numerical} means have been unsuccessful:
\begin{itemize}
\item no evidence for the existence of \emph{four} MU bases in $\mathbb{C}^6$ has been found \cite{butterley+07};
\item strong numerical evidence against the existence of various MU \emph{constellations} (corresponding to \emph{subsets} of four MU bases) has been obtained, making the existence of a complete set highly unlikely \cite{brierley+08}. 
\end{itemize} 
Some rigorous results have been obtained by restricting the search to MU bases of a specific form:  
\begin{itemize}
\item selecting a first Hadamard matrix and then searching for MU vectors with components given by suitable roots of unity leads to no more than two MU complex Hadamard matrices, or three MU bases in $\mathbb{C}^6$ \cite{bengtsson+07};
\item Grassl \cite{grassl04} has shown that only \emph{finitely} many vectors exist which are MU with respect to the identity and a given complex Hadamard matrix related to the Heisenberg-Weyl group in $\mathbb{C}^6$. Again, no more than two MU Hadamard matrices emerge, giving rise to at most three MU bases; it is thus impossible to base the construction of a complete set on the Heisenberg-Weyl group.
\end{itemize}

The strategy of this paper will be to generalize Grassl's approach by removing the restriction that the second MU basis be related to the Heisenberg-Weyl group. Instead, we will consider many different choices for the second MU basis, thoroughly sampling the set of currently known complex Hadamard matrices in $\mathbb{C}^6$. We will find that none of the matrices studied can be used to construct a complete set of MU bases. Taken together, these negative instances provide further strong support for the conjecture that no seven MU bases exist in dimension six.  

Let us now present the outline of our argument. In Sec. \ref{knownHadamards}, we briefly describe the set of known complex Hadamard matrices in dimension six. Then, we explain in Sec. \ref{construction} how to construct all vectors that are MU with respect to both the standard basis of $\mathbb{C}^6$ and a second basis, defined by an arbitrary fixed Hadamard matrix. We illustrate the algorithm for $d=3$ only to  rediscover the known complete set of four MU bases. Then, whilst rederiving Grassl's result for $d=6$, we will explain the subtle interplay between algebraic and numerical calculations in this approach. Sec. \ref{Results} presents our findings which we obtain by applying the algorithm to nearly 6000 Hadamard matrices of dimension six. Conclusions are drawn in the final section.

\section{Complex Hadamard matrices in dimension six}

\label{knownHadamards}

Traditionally, a Hadamard matrix $H$ in dimension $d$ is understood to have elements $\pm 1$ only and to satisfy the condition $H^\dagger H = d I$, where $I$ is the identity. In the context of MU bases, it is customary to call $H$ a \emph{Hadamard} matrix if it is unitary and its matrix elements are of the form 
\begin{equation} \label{moduli}
\left\vert H_{ij}\right\vert 
 =\frac{1}{\sqrt{d}}\, , \qquad i,j = 0,1, \ldots , d-1 \, . 
\end{equation}
The $d$ vectors formed by the columns of such a matrix provide an orthonormal  basis of $\mathbb{C}^d$. Each of these vectors is mutually unbiased with respect to the standard basis, naturally associated with the identity matrix $I$. It is convenient to identify a Hadamard matrix with the MU basis formed by its columns.

Two Hadamard matrices are \emph{equivalent} to each other, $H^{\prime } \approx H$, if one can be obtained from the other by permutations of its columns and its rows, and by the multiplication of its columns and rows with individual phase factors. Explicitly, the equivalence relation reads 
\begin{equation} \label{equivalence}
H^{\prime } = M_{1}HM_{2} \, ,
\end{equation}
where $M_1$ and $M_2$ are \emph{monomial} matrices, i.e. they are unitary and have only one nonzero element in each row and column. Consequently, each Hadamard matrix is equivalent to a \emph{dephased} Hadamard matrix, the first row and column of which have entries $1/\sqrt{d}$ only.  

All (complex) Hadamard matrices are known for dimensions $d \leq 5$ but there is no exhaustive classification for $d = 6$. It is useful to briefly describe the Hadamard matrices known to exist in dimension six since we will `parametrize' the search for MU bases in terms of Hadamard matrices. We use the notation introduced in \cite{tadej+06} the authors of which maintain  an online catalog of Hadamard matrices \cite{Hadamardsonline}.

Each point in Fig. \ref{AllHadamards}, an updated version of a figure presented in \cite{bengtsson+07}, corresponds to one Hadamard matrix of dimension six, except for the interior of the upper circle where a point represents two Hadamard matrices (cf. below). There is one \emph{isolated point}, representing the spectral matrix $S$ given in \cite{moorhouse01}, also known as Tao's matrix \cite{tao04}. Three sets of Hadamard matrices labeled by a \emph{single parameter} are known: the Di\c{t}\u{a} family $D(x)$ introduced in \cite{dita04}, a family of symmetric  matrices denoted by $M(t)$ \cite{matolcsi+07} and the family of all Hermitean Hadamard matrices $B(\theta)$ \cite{beauchamp+06}. Two \emph{two-parameter} families of Hadamard matrices are known to arise from discrete Fourier-type transformations $F(x_1,x_2)$ in $\mathbb{C}^6$, and from their transpositions, $F^T(x_1,x_2)$ \cite{haagerup96}. The Sz\"{o}ll\H{o}si family $X(a,b)$  is the only other known two-parameter set \cite{szollosi08}. Interestingly, the matrix $X(0,0)$ can be shown to be equivalent to $F(1/6,0)$, and there is a second possibility to define a matrix at this point, giving rise to $X^T(0,0) \approx F^T(1/6,0)$ \cite{bengtsson08}. We have noticed that such a doubling actually occurs for \emph{all} values of the parameters $(a,b)$ leading to a set of Hadamard matrices $X^T(a,b)$ inequivalent to $X(a,b)$. Hence, the interior of the upper circle in Fig. \ref{AllHadamards} represents two layers of Hadamard matrices which are glued together at its boundary. Topologically, the Sz\"{o}ll\H{o}si family $X(a,b)$ and the set $X^T(a,b)$ thus combine to form the surface of a sphere.  Appendix \ref{defnHadamards} lists the explicit forms of Hadamard matrices as well as the parameter ranges which have been reduced to their \emph{fundamental regions} using the equivalence relation ({\ref{equivalence}). 

Fig. \ref{AllHadamards} also shows \emph{equivalences} between Hadamard matrices simultaneously belonging to different families. The circulant Hadamard matrix $C$ \cite{bjorck+95}, for example, embeds into the Hermitean family which in turn is given by the boundary of the Sz\"{o}ll\H{o}si families; interestingly, the Di\c{t}\u{a} matrices are also contained therein. Lining up some of the points where different families overlap suggests that we arrange the Hadamard matrices in a symmetrical way. Then,  a reflection about the line passing through the points $F(0,0)$ and $S$ maps $H(\mathbf{x})$ to $H(\mathbf{-x})$ if $H(\mathbf{x})$ is a member of the Di\c{t}\u{a}, Hermitean or symmetric families; furthermore, the same reflection sends $H(\mathbf{x})$ to  $H^{T}(\mathbf{x})$ if the matrix $H(\mathbf{x})$ is taken from Di\c{t}\u{a}, Hermitean or Fourier families. For the Sz\"{o}ll\H{o}si family, the reflection about the vertical axis must be supplemented by a change of layer in order to get from $X(a,b)$ to $X^T(a,b)$. We will see that the findings presented in Sec. \ref{Results} echo this symmetry  which we will explain in the conclusion.   

Let us finally mention that the known families of Hadamard matrices come in two different types, \emph{affine} and \emph{non-affine} ones. The set $H(\mathbf{x})$ is affine if it can be written in the form
\begin{equation} \label{defAffine}
H(\mathbf{x})=H(0)\circ \text{Exp}[R(\mathbf{x})]
\end{equation}
for some matrix $R$; the open circle denotes the Hadamard (elementwise) product of two matrices, $(A\circ B)_{ij}=A_{ij}B_{ij}$, and $\text{Exp}[R]$ represents the matrix $R$ elementwise exponentiated: $(\text{Exp}[R])_{ij} = \exp R_{ij}$. Both Fourier-type families and the Di\c{t}\u{a} matrices are affine (cf. Appendix \ref{defnHadamards}) while the symmetric, Hermitean and Sz\"{o}ll\H{o}si families are not.

\section{Constructing MU Vectors} \label{construction}

In this section, we make explicit the conditions on a vector $\mathbb{C}^6$ to be MU with respect to the standard matrix and a fixed Hadamard matrix, i.e. to the pair $\{I,H\}$. Then we outline an algorithm to construct \emph{all} solutions of the resulting multivariate polynomial equations, allowing us to check how many additional MU Hadamard matrices do exist. We illustrate this approach by constructing a complete set of \emph{four} MU bases in dimension $d=3$, and we reproduce Grassl's result for $d=6$ in order to explain that this approach produces rigorous results in spite of inevitable numerical approximations.

\subsection{MU vectors and multivariate polynomial equations}

A vector $\ket{v} \in \mathbb{C}^d$ is MU with respect to the standard basis (associated with the columns of the identity $I$) if each of its components has modulus $1/\sqrt{d}$. Furthermore, $\ket{v}$ is MU with respect to a fixed Hadamard matrix $H$ if $|\langle h(k)|v \rangle |^{2} = 1/d$, where $\ket{h(k)}$ is the state associated with the $k^{\mbox{\footnotesize{th}}}$ column $h(k)$ of $H$, $k=0,\ldots,d-1$. 

Let us express these conditions on $\ket{v}$ in terms of its components $v_j$, written as 
\begin{equation} \label{para vec}
\sqrt{d}v_{j}=\left\{ 
\begin{array}{ll}
           1 & \quad j = 0 \, ,  \\ 
x_{j}+iy_{j} & \quad j = 1, \ldots, d-1\, ,
\end{array}
\right. 
\end{equation}
where $x_{j},y_{j}$ are $2(d-1)$ real parameters. The overall phase of the state $\ket{v}$ is irrelevant which allows us to fix the phase of its first component.
Then, the first set of constraints on the state $\ket{v}$ reads 
\begin{equation} \label{modulus1}
x_{j}^{2}+y_{j}^{2}=1 \, , \quad j = 1, \ldots, d-1\, , 
\end{equation}
and the second set is given by 
\begin{equation} \label{MU in components}
\left|\sum_{j=0}^{d-1} h^*_j(k) v_j \right|^2 
    \equiv \left|\sum_{j=0}^{d-1} H^\dagger_{kj} (x_j+iy_j)\right|^2 
    = \frac{1}{d} \, , \quad k=0, \ldots, d-2 \, ,
\end{equation}
where the state $\ket{h(k)}$ has components $h_j(k) \equiv H_{jk}$, $0=1,\ldots,d-1$.
The completeness relation of the orthonormal basis $\{\ket{h(k)}$, $k=0, \ldots, d-1\}$,  
implies that if a state $\ket{v}$ is MU with respect to $(d-1)$ of its members, it is also MU with respect to the remaining one. Therefore, it is not necessary to include $k\equiv d-1$ in Eqs. (\ref{MU in components}).  

For each given Hadamard matrix $H$, Eqs. (\ref{modulus1}) and (\ref{MU in components}) represent $2(d-1)$ simultaneous coupled quadratic equations for $2(d-1)$ real variables. Once we know \emph{all} solutions of these equations, we know \emph{all} vectors $\ket{v}$ MU with respect to the chosen pair of bases $\{I,H\}$.
Analysing the set of solutions will reveal whether they form additional MU Hadamard matrices, or, equivalently, MU bases.  

If Eqs. (\ref{modulus1}) and (\ref{MU in components}) were \emph{linear}, one could apply Gaussian elimination to bring them into `triangular' form. The resulting  equations would have the same solutions as the original ones but the solutions could be obtained easily by successively solving for the unknowns.

The solutions of Eqs. (\ref{modulus1}) and (\ref{MU in components}) can be found using \emph{Buchberger's algorithm} \cite{buchberger65} which generalizes Gaussian elimination to \emph{(nonlinear) multivariate polynomial} equations. In this approach, a set of polynomials ${\cal P} \equiv \{p_{n}(\mathbf{x}), n=1, \ldots, N\}$ is transformed into a different set of polynomials ${\cal G} \equiv \{ g_{m}(\mathbf{x}), m=1, \ldots, M\}$ (usually with $ M\neq N$) such that the equations ${\cal P}=0$ and ${\cal G}=0$ possess the \emph{same} solutions; here ${\cal P}=0$ is short for $p_{n}(\mathbf{x}) = 0, n=1, \ldots, N$. Technically, one constructs a \emph{Gr\"{o}bner basis} ${\cal G}$ of the polynomials ${\cal P}$ which requires a choice of variable ordering \cite{buchberger65}. The transformed equations ${\cal G}=0$ will be straightforward to solve due their `triangular' form:  one can find all possible values of a first unknown by solving for the zeros of a polynomial in a \emph{single} variable; using each of these solutions will reduce one or more of the remaining equations to single-variable polynomials, allowing one to solve for a second unknown, etc. This process iteratively generates all solutions of ${\cal G}=0$ and, therefore, all solutions of the original set of equations, ${\cal P}=0$.

A Gr\"{o}bner basis exists for any set of polynomial equations with a finite number of variables. However, the number of steps required to construct a Gr\"{o}bner basis tends to be large even for polynomials of low degrees and a small number of unknowns. Thus, Buchberger's algorithm is most conveniently applied by means of algebraic software programs. We have used the implementation \cite{salsa} of this algorithm suitable for the computational algebra system Maple \cite{Maple} since we found it to be particularly fast for the system of equations under study. 

Let us now make explicit how to construct all vectors MU with respect to a pair $\{ I, H\}$ by solving the multivariate polynomial equations (\ref{modulus1}) and (\ref{MU in components}) using Buchberger's algorithm. We will consider two cases in dimensions $d=3$ and $d=6$, respectively, which have been solved before but they are suitable to illustrate the construction and to discuss some of its subleties. 

\subsection{Four MU bases in $\mathbb{C}^3$}

In dimension $d=3$, four MU bases are known to exist. We will now show how to construct two MU Hadamard matrices $H_2$ and $H_3$ given a pair $\{I,H\}$. The resulting three MU Hadamard matrices plus the identity provide a complete set of four MU bases in $\mathbb{C}^3$.

\paragraph{1. Choose a Hadamard}

In dimension three, all Hadamard matrices are known and there is only one choice for a dephased Hadamard matrix \cite{haagerup96} given by the Fourier matrix, 
\begin{equation} \label{F3}
F_{3}=\frac{1}{\sqrt{3}}\left( 
\begin{array}{ccc}
1 & 1 & 1 \\ 
1 & \omega & \omega ^{2} \\ 
1 & \omega ^{2} & \omega%
\end{array}
\right) \, ,
\end{equation}
where $\omega =\exp (2\pi i/3)$ is a third root of unity. 

\paragraph{2. List the constraints}

We want to find all states $\ket{v} \in \mathbb{C}^3$ which are MU with respect to the columns of the identity matrix $I$ and the Fourier matrix $F_3$. Using the four real parameters $x_1,x_2,y_1$, and $y_2$ introduced in (\ref{para vec}), the constraints (\ref{modulus1}) and (\ref{MU in components}) read explicitly 
\begin{eqnarray}
                        1- x_{1}^{2}-y_{1}^{2} &=&0 \, , \nonumber\\
                        1- x_{2}^{2}-y_{2}^{2} &=&0 \, ,  \nonumber \\
               x_{1}+x_{2}+x_{1}x_{2}+y_{1}y_{2} &=&0 \, , \nonumber \\
x_{1}+x_{2}-\sqrt{3}y_{1}+\sqrt{3}y_{2}+x_{1}x_{2}
-\sqrt{3}x_{1}y_{2}+\sqrt{3}y_{1}x_{2}+y_{1}y_{2} &=&0 \, .  \label{dim3eqs}
\end{eqnarray}
The solutions of these four coupled quadratic equations in four real variables, ${\cal P}=0$, will tell us whether additional Hadamard matrices exist which are MU with respect to the Fourier matrix $F_3$. 

\paragraph{3. Construct the solutions}

By running Buchberger's algorithm, we find the Gr\"{o}bner basis ${\cal G}$ associated with the polynomials in Eqs. (\ref{dim3eqs}). Equating the resulting four polynomials $g_n({\bf x}), n=1, \ldots, 4$, to zero, gives rise to the equations
\begin{eqnarray}
                   3y_{2}-4y_{2}^{3} &=& 0 \, , \nonumber\\
                  1-x_{2}-2y_{2}^{2} &=& 0 \, , \nonumber\\
     1+2x_{1}+4y_{1}y_{2}-4y_{2}^{2} &=& 0 \, , \nonumber\\
 3-4y_{1}^{2}+4y_{1}y_{2}-4y_{2}^{2} &=& 0 \, . \label{GB1-4}     
\end{eqnarray}
This set is `triangular' in the sense that solutions can be found by iteratively determining the roots of polynomials for single variables only. The first equation has three solutions,
\begin{equation}
y_{2} \in \{ 0 \, , \pm \sqrt{3}/2\}\, ;
\end{equation}
next, the second equation implies that 
\begin{equation}
x_2 = \left\{ 
\begin{array}{ll}
           0 & \quad \mbox{if } y_2 = 0 \, ,  \\ 
           2 & \quad \mbox{if } y_2 = \pm \sqrt{3}/2\, ;
\end{array}
\right.
\end{equation}
etc. Altogether, there are six solutions,
\begin{equation}
\begin{array}{ll}
\mathbf{s}_a  = \frac{1}{2}(-1,-1,\sqrt{3},\sqrt{3})\, , \quad &
\mathbf{s}_b  = \frac{1}{2}(-1,2,-\sqrt{3},0)\, , \nonumber \\
\mathbf{s}_c  = \frac{1}{2}(2,-1,0,-\sqrt{3}) \, , \quad &
\mathbf{s}_d  = \frac{1}{2}(-1,-1,-\sqrt{3},-\sqrt{3}) \, , \nonumber \\
\mathbf{s}_e  = \frac{1}{2}(2,-1,0,\sqrt{3}) \, , \quad &
\mathbf{s}_f  = \frac{1}{2}(-1,2,\sqrt{3},0)  \,  , 
\end{array}
\end{equation}
defining $\mathbf{s}=(x_{1},x_{2},y_{1},y_{2})$. 

Since the degrees of the polynomials ${\cal G}$ in Eqs. (\ref{GB1-4}) do not exceed three, we are able to obtain analytic expressions for its solutions. This, however, is a fortunate coincidence due to the simplicity of the problem: in general, we will need to determine the roots of higher-order polynomials (cf. the example presented in Sec. \ref{fourierind=6}) which requires numerical methods. The resulting complications will be discussed in Sec. \ref{numeric approx}.

\paragraph{4. List all MU vectors}

Upon substituting the solutions $\mathbf{s}_a$ to $\mathbf{s}_f$ into (\ref{para vec}), one obtains six vectors 
\begin{equation} \label{sixvectors}
\begin{array}{l}
v_{a} =\frac{1}{\sqrt{3}}\left( 
\begin{array}{c}
1 \\ 
\omega \\ 
\omega%
\end{array}%
\right) \, , \quad 
v_{b}=\frac{1}{\sqrt{3}}\left( 
\begin{array}{c}
1 \\ 
\omega^2 \\ 
1%
\end{array}%
\right) \, , \quad  
v_{c}=\frac{1}{\sqrt{3}}\left( 
\begin{array}{c}
1 \\ 
1 \\ 
\omega^2%
\end{array}%
\right) \, , \\
v_{d}=\frac{1}{\sqrt{3}}\left( 
\begin{array}{c}
1 \\ 
\omega^2 \\ 
\omega^2%
\end{array}%
\right) \, , \quad 
v_{e} =\frac{1}{\sqrt{3}}\left( 
\begin{array}{c}
1 \\ 
1 \\ 
\omega%
\end{array}%
\right) \, , \quad  
v_{f}=\frac{1}{\sqrt{3}}\left( 
\begin{array}{c}
1 \\ 
\omega \\ 
1 %
\end{array}%
\right) \, ,
\end{array}
\end{equation}
which are MU with respect to the columns of both the matrices $I$ and $F_{3}$. No other vectors with this property exist, leaving us with $v_a, \ldots, v_f$, as the only candidates for the columns of additional MU Hadamard matrices. 

\paragraph{5. Analyse the vectors}

The six vectors in (\ref{sixvectors}) allow us to define an additional Ha\-da\-mard matrix only if any three of them are orthogonal; for a second Hadamard matrix the remaining three must be orthogonal among themselves \emph{and} MU to the first three. Calculating the inner products between all pairs of the vectors $v_a$ to $v_f$ shows that they indeed fall into two groups with the required properties. Consequently, we have constructed a complete set of four MU bases in $\mathbb{C}^3$, corresponding to the set $\{I,F_{3},H_{2},H_{3}\}$ where the columns of the matrices $H_2$ and $H_3$ are given by $\{v_a,v_b,v_c\}$ and $\{v_d,v_e,v_f\}$, respectively. 

We have also checked that the construction procedure works in dimensions $d=2,5$ and $d=7$ where it correctly generates complete sets of $(d+1)$ MU bases. The matrices $F_{2}$, $F_{3}$ and $F_{5}$ are the \emph{only} dephased Hadamard matrices in dimensions $d=2,3$ and $d=5$, and there is only one way to construct complete MU bases from the vectors obtained. We have thus shown that the Heisenberg-Weyl construction of a complete set of MU bases is essentially \emph{unique} in dimensions two, three and five, correctly reproducing known results \cite{boykin+07,kostrikin+94}.

\subsection{Three MU bases in $\mathbb{C}^6$} \label{fourierind=6}

In $d=6$, the existence of seven MU bases is an open problem. We will search for all states $\ket{v}$ which are MU with respect to the identity $I$ and the six-dimensional equivalent of $F_3$ given in (\ref{F3}), the dephased Fourier matrix 
\begin{equation} \label{F6}
F_6 = \left( 
\begin{array}{cccccc}
1 & 1 & 1 & 1 & 1 & 1 \\ 
1 & \omega & \omega^{2} & \omega^{3} & \omega^{4} & \omega^{5} \\ 
1 & \omega^{2} & \omega^{4} & 1 & \omega^{2} & \omega^{4} \\ 
1 & \omega^{3} & 1 & \omega^{3} & 1 & \omega^{3} \\ 
1 & \omega^{4} & \omega^{2} & 1 & \omega^{4} & \omega^{2} \\ 
1 & \omega^{5} & \omega^{4} & \omega^{3} & \omega^{2} & \omega%
\end{array}
\right) ,
\end{equation}
with $\omega = \exp (\pi i/3)$ now being the sixth root of unity. This problem has been studied in the context of biunimodular sequences \cite{bjorck+95} and in relation to MU bases \cite{grassl04}. It is impossible to complement the pair $\{I,F_6\}$ by more than one Hadamard matrix MU with respect to $F_6$. Thus, the construction method of MU bases in prime-power dimensions which is based on the Heisenberg-Weyl group, has no equivalent in the composite dimension $d=6$. We will now reproduce this negative result.    

Having chosen the first Hadamard matrix to be $F_6$, we can write down the conditions which the components of a state $\ket{v}$ must satisfy, ${\cal P}=0$. After some algebraic operations detailed in Appendix \ref{SimplifyEqns}, one obtains the equations
\begin{eqnarray} 
x_{{1}}+x_{{5}}+x_{{1}}x_{{2}}+x_{{2}}x_{{3}}+x_{{3}}x_{{4}}+x_{{4}}x_{{5}}   
  +y_{{1}}y_{{2}}+y_{{2}}y_{{3}}+y_{{3}}y_{{4}}+y_{{4}}y_{{5}} &=& 0 \, , \nonumber \\
y_{{1}}-y_{{5}}+x_{{1}}y_{{2}}-x_{{2}}y_{{1}}+x_{{2}}y_{{3}}-x_{{3}}y_{{2}}    
  +x_{{3}}y_{{4}}-x_{{4}}y_{{3}}+x_{{4}}y_{{5}}-x_{{5}}y_{{4}} &=& 0 \, , \nonumber \\
x_{{3}}+x_{{1}}x_{{4}}+x_{{2}}x_{{5}}+y_{{1}}y_{{4}}+y_{{2}}y_{{5}} &=& 0 \, ,\nonumber\\
x_{{2}}+x_{{4}}+x_{{1}}x_{{3}}+x_{{1}}x_{{5}}+x_{{2}}x_{{4}}+x_{{3}}x_{{5}}  
   +y_{{1}}y_{{3}}+y_{{1}}y_{{5}}+y_{{2}}y_{{4}}+y_{{3}}y_{{5}} &=& 0 \, , \nonumber\\
y_{{2}}-y_{{4}}+x_{{1}}y_{{3}}-x_{{1}}y_{{5}}+x_{{2}}y_{{4}}-x_{{3}}y_{{1}} 
  +x_{{3}}y_{{5}}-x_{{4}}y_{{2}}+x_{{5}}y_{{1}}-x_{{5}}y_{{3}} &=& 0 \, , \label{F00}
\end{eqnarray}
which must be supplemented by the five conditions (\ref{modulus1}) arising for $d=6$. 

We need to find all solutions of these ten coupled equations ${\cal P}=0$ which are quadratic in ten real variables. The Gr\"{o}bner basis ${\cal G}$ associated with the set ${\cal P}$ consists of $36$ polynomials of considerably higher degrees. We reproduce only the first one of the new set of equations, ${\cal G}=0$, 
\begin{eqnarray} \label{y5}
-245025\,y_{{5}}+4318758\,{y_{{5}}}^{3}-28135161\,{y_{{5}}}^{5}+89685000\,{%
y_{{5}}}^{7}-158611892\,{y_{{5}}}^{9} & & \nonumber \\
+177275680\,{y_{{5}}}^{11}-150745472\,{y_{{5}}}^{13}+104333824\,{y_{{5}}}%
^{15}-43667456\,{y_{{5}}}^{17} & & \\
+2351104\,{y_{{5}}}^{19}+4882432\,{y_{{5}}}^{21}-1703936\,{y_{{5}}}%
^{23}+262144\,{y_{{5}}}^{25} & = & 0 \, , \nonumber
\end{eqnarray}
being of order $25$ in the single variable $y_5$. This equation admits $15$ real solutions, 
\begin{equation} \label{y5solutions}
y_5 \in \{ 0,\pm 1,\pm \frac{1}{2},\pm \frac{\sqrt{3}}{2},\pm \frac{1}{2}(1+\sqrt{3}
),\pm \frac{1}{2}(1-\sqrt{3}),\pm 0.988940\ldots ,\pm 0.622915\ldots \} \, ,
\end{equation}
the last four of which we only find numerically. Due to the triangular structure resulting from Buchberger's algorithm, there will be equations (at least one) containing only $y_5$ and one other single variable. For each value of $y_5$ taken from (\ref{y5solutions}), they reduce to single-variable polynomials the roots of which can be determined to desired numerical accuracy; etc. Keeping track of all possible branches we obtain $48$ vectors that satisfy the Eqs. (\ref{F00}). 

Having determined the candidates for columns of MU Hadamard matrices, we calculate the inner products among all pairs of the 48 vectors. It turns out that there are 16 different ways to group them into bases of $\mathbb{C}^{6}$. However, no two of these bases are MU with respect to each other. Consequently, it is possible to form at most 16 different \emph{triples} of MU bases which include $F_6$. It also follows that the Fourier matrix $F_6$ (or any other unitarily equivalent element of the Heisenberg-Weyl group \cite{grassl04}) \emph{cannot} be supplemented by two MU Hadamard matrices--no four MU bases can exist.

There are, however, many choices other than $H=F_6$ for a dephased Hadamard matrix  in dimension six. In Sec. \ref{Results}, we will repeat the calculations just presented for a large sample of currently known Hadamard matrices. Before doing so, we will discuss the fact that we are able to construct the desired vectors only approximately.  In the following section we show that sufficiently high numerical accuracy allows us to draw \emph{rigorous} conclusions about the properties of the exact vectors. 

\subsection{The Impact of Numerical Approximations} \label{numeric approx}

The previous section illustrated that the problem of finding MU vectors with respect to the identity $I$ and a given Hadamard matrix $H$ can be reduced to successively solving for the roots of polynomials of a single variable. These roots, however, can only be found approximately. Does the approximation prevent us from drawing rigorous conclusions about the properties of the MU vectors we construct? We will argue now that it remains possible to find upper bounds on the number of MU vectors with the desired properties. 

Consider the system of polynomials ${\cal P}=\{p_{n}(\mathbf{x}), n=1, \ldots, 10\}$ in the variables $\mathbf{x\in }\mathbb{R}^{10}$ resulting from some chosen Hadamard matrix $H$, and calculate a Gr\"{o}bner basis, ${\cal G}=\{g_{m}(\mathbf{x}), m=1, \ldots, M \}$. The roots of the equations ${\cal P}=0$  and ${\cal G}=0$ are identical by construction. Since ${\cal G}=0$ corresponds to a `triangular' set, its roots can be found iteratively but, in general, no closed form will exist. The implementation of Buchberger's algorithm which we have chosen finds these roots with user-specified accuracy, relying on the theory presented in \cite{rouillier99}. 

Suppose that ${\cal G}=0$ has two roots $\mathbf{s}_a$ and $\mathbf{s}_b$, to which we have found approximations, $\mathbf{s}_A$ and $\mathbf{s}_B$. The associated approximate exact states, $\ket{v_a}$ and $\ket{v_b}$, differ from the approximate states, $\ket{v_A}$ and $\ket{v_B}$, by error terms $\ket{\delta v_a} = \ket{v_A} - \ket{v_a}$ and similarly for the second solution. The components of the vectors $\ket{\delta v_a}$ all have moduli smaller than the user-defined accuracy of $10^{-r}$, say. If the inner product of the exact states $\ket{v_a}$ and $\ket{v_b}$ has a non-zero modulus, $\Delta>0$, then they are \emph{not} orthogonal. We can detect this by calculating the inner product of the \emph{approximate} states,
\begin{eqnarray} \label{estimate}
|\bra{v_A}v_B \rangle|
	&=&|\bra{v_a}v_b\rangle+\bra{v_a}\delta v_b\rangle+\bra{\delta v_a}v_b\rangle
	       +\bra{\delta v_a}\delta v_b\rangle| \nonumber\\
	&\leq &|\bra{v_a}v_b\rangle|+|\bra{v_a}\delta v_b\rangle|+|\bra{\delta v_a}v_b\rangle|
	       +{\cal{O}} (10^{-2r}) \nonumber \\
	&\leq & |\bra{v_a}v_b\rangle|+10\sqrt{2} \times 10^{-r}+{\cal{O}} (10^{-2r}) \, ,
\end{eqnarray}
using $||\ket{\delta v_a}|| \leq 5\sqrt{2} \times 10^{-r}$ and $||\ket{ v_a}||=1$. Thus, a non-zero lower bound for the exact scalar product follows if the approximate inner product is \emph{larger} than $\sqrt{2} \times 10^{-r+1}$. In other words, we may conclude that the exact states are non-orthogonal if we ensure that the error in the approximate scalar product is negligible, i.e. $\Delta \geq |\bra{v_A}v_B \rangle|-\sqrt{2} \times 10^{-r+1} >0$. A similar argument allows us to exclude that two approximate states are MU with respect to each other. 

We determine the roots of ${\cal G}=0$ to $r=20$ significant digits which proves sufficient to put relevant limits on the properties of the vectors constructed in dimension six.  The results presented in the Sections \ref{specialmatrices} and \ref{affinefamilies} thus represent rigorous limits on the number of vectors MU with respect to specific Hadamard matrices and hence on the number of MU bases.

\section{Constructing MU Bases in Dimension Six}\label{Results}

We are now in a position to present the main results of this paper.
We will consider one Hadamard matrix $H$ at a time constructing all additional Hadamard matrices MU with respect to the chosen one. Picking matrices both systematically and randomly, we will find that not a single one is compatible with the existence of four MU bases.

More specifically, we will determine two quantities for each chosen Hadamard matrix $H$.  The number $N_{v}$ equals the number of vectors MU with the pair $\{I,H\}$, and the number $N_t$ provides an upper bound on how many different triples of MU bases $\{I,H,H^\prime\}$ exist.

\subsection{Special Hadamard Matrices} \label{specialmatrices}

To begin, we consider the Hadamard matrices on the symmetry axis of Fig. (\ref{AllHadamards}): the Fourier matrix $F_6\equiv F(0,0)$ being invariant under transposition, the Di\c{t}\u{a} matrix $D_0 \equiv D(0)$ which is both symmetric and Hermitean, the circulant matrix $C$, and the Spectral matrix $S$. These matrices are special in the sense that they are either isolated or belong to different Hadamard families simultaneously.

The first row of Table \ref{res:special} completes the findings of Sec. \ref{fourierind=6} obtained for the \emph{Fourier} matrix $F_6$: there are $N_v=48$ vectors MU with respect to both $I$ and $F_6$ that can be arranged in  $N_t=16$ different ways to form a second Hadamard matrix $H^\prime$ being MU with respect to $F_6$. However, no two of these 16 Hadamard matrices are MU between themselves, limiting the number of MU bases containing $F_6$ to three.

A similar analysis for the \emph{Di\c{t}\u{a}} matrix $D_0$ reveals that there are 120 vectors MU to its columns and those of the identity, 60 of which form ten bases but none of these are MU with respect to each other. Whilst ten triples of MU bases exist, sets of four MU bases which include $D_0$ do \emph{not} exist. 

Interestingly, the components of the 120 vectors have phases $\phi$ which take values in a small set only,
\begin{equation}
\phi_{D} \equiv \{0,\pi ,\pm \pi/12,\ldots ,\pm 11\pi/12,\pm  \alpha \} \, ,
\end{equation}
where $\tan \alpha =2.$ This result\footnote{Such a restricted set of  phases also occurs for other members of the Di\c{t}\u{a} family. For example, all 48 vectors MU with the pair $\{I,D(1/8)\}$ have phases limited to the set
$\phi_{D} \cup \{\pm \beta \} $ where $\tan \beta =3$.} agrees with the one obtained by Bengtsson et al. \cite{bengtsson+07} (note, however, that the descriptions given in the last two entries of the list in their Sec. 7 must be swapped). What is more, our approach \emph{proves} that these authors have been able to identify \emph{all} vectors MU with the pair $\{I,D_0\}$ by means of their ansatz for the form of MU vectors. In fact, the value of $N_{t}$ in Table \ref{res:special} given for $D_0$ is \emph{exact}, not an upper bound since the phases of the MU states $\ket{v}$ are known in closed form. 

The \emph{circulant} matrix $C$ permits 56 MU vectors, which can be arranged into 4 different bases, $N_t=4$.

The \emph{spectral} matrix $S$ is the only known \emph{isolated} Hadamard matrix. We find $90$ MU vectors but not a single sextuple of orthonormal ones among them. Thus, the pair $\{I,S\}$ cannot even be extended to a triple of MU bases. 

\subsection{Affine Families} \label{affinefamilies}

Table \ref{res:affine} collects the properties of vectors MU with respect to the pair $\{I,H\}$ where $H$ is an affine Hadamard matrix, i.e. taken either from the one-parameter set discovered by Di\c{t}\u{a} or from the two-parameter Fourier families. Again, we have sampled the relevant parameter spaces both systematically and randomly. 

The set of \emph{Di\c{t}\u{a}} matrices $D(x)$ depends on a single continuous parameter $x$, with $|x| \leq  1/8$. We have sampled the interval in steps of size $1/144$ making sure that the resulting grid of points include the $24$th roots of unity which play an important role for $D_0$, so 
\begin{equation} \label{Ditagrid}
\Gamma_{D} =\{a/144:a=\pm 1,\pm 2\ldots , \pm 18\} \, ;
\end{equation}
note that the matrix $D_0$ has been left out. The number of vectors MU with the pair $\{I,D(x)\}$ depends on the \emph{value} of the parameter $x$: the Di\c{t}\u{a} matrices $D(x)$ on the grid $\Gamma_D$ allow for 48, 72 or 120 MU vectors which can be grouped into into four additional Hadamard matrices. Since they are not MU between themselves, there are at most three MU bases containing any of these Di\c{t}\u{a} matrices. 

The results obtained from \emph{randomly} picking points in the fundamental interval are in line with the observations made for grid points. Fig. \ref{Dita fig} shows $N_v$, the number of vectors MU with respect to the pair $\{ I, D(x) \}$ for all 536 values of the parameter $x$ which we have considered. The function $N_v(x)$ appears to be symmetric about $x=0$ and piecewise constant, dropping from 120 for small values of $x$ to 72 at $x \simeq \pm 0.0177$, and to 48 at the end points of the interval, $x=\pm 1/8$. The values for $N_v$ can be found in Table \ref{res:affine}.

The results for members of \emph{Fourier} family $F({\bf x})$ are qualitatively similar. Picking values of ${\bf x} \equiv (x_1,x_2)$ either randomly in the fundamental area or from the two-dimensional grid
\begin{equation} \label{Fouriergrid}
\Gamma_{F} =\{(a,b)/144 : a=1,2,\ldots ,24,\, b=0,1,\ldots ,12, \, a\geq 2b\} \, ,
\end{equation}
invariably leads to 48 vectors being MU to the columns of the pair
$\{ I, F({\bf x})\}$. There are eight different ways to form additional Hadamard matrices for each point considered except for the matrix $F(1/6,0)$  with an upper bound of 70 triples. It is important to realize that Grassl's result---the construction of complete sets of MU bases cannot be based on the Heisenberg-Weyl group in dimension 
$d=6$---also holds for the 2,168 other Fourier matrices we have considered.

The situation is similar when turning to the family of \emph{transposed Fourier} matrices, $F^T({\bf x})$. The number $N_v$ equals 48 throughout and a second Hadamard matrix can be formed in eight different ways, and only   matrix $F^T(1/6,0)$ allows for 70 different triples, eight being the norm. 

\subsection{Non-Affine Families} \label{nonaffinefamilies}

The equations ${\cal P}=0$ encoding MU vectors for the symmetric $M(t)$, Hermitean $B(\theta)$ and Sz\"oll\H{o}si $X(a,b)$ families turn out to be more challenging from a computational perspective: the program has, in general,  not been able to construct the associated Gr\"obner bases ${\cal G}$. The problem is not a fundamental one---the desired Gr\"{o}bner bases do exist but it appears that their construction requires more memory than the 16GB available to us.

We suspect that the difficulties are due to the fact that, for non-affine matrices, the coefficients of the polynomials ${\cal P}=0$ are no longer equal to fractions or simple roots of integers.  When approximating the coefficients in question by fractions we obtain different sets of polynomials, $\tilde {\cal P}$, and the program indeed succeeds in constructing the corresponding Gr\"obner bases, $\tilde {\cal G}$, outputting (approximate) MU vectors $\ket{\tilde v}$. Being continuous functions of the coefficients, the approximate vectors will resemble the exact ones, $\ket{\tilde v} \simeq \ket{v}$. However, the \emph{number} of MU vectors may change discontinuously if $\tilde {\cal P}=0$ is considered instead of ${\cal P}=0$, similar to the discontinuous change in the number $N_v$ for the family $D(x)$ near $x \simeq 0.0177$, shown in Fig. \ref{Dita fig}. In other words, it could happen that we `lose' some solutions due to a geometric instability as a consequence of modifying the defining polynomials. 

To determine the impact of such an approximation, we have studied how the number $N_v$ of MU vectors changes in a case for which we know rigorous bounds. We retain only five significant digits of the coefficients in the equations ${\cal P}=0$ associated with the family $D(t)$ and solve for the approximate MU vectors. The inset of Fig. \ref{Dita fig} shows that the plateaus of $120$ and $72$ MU vectors continue to be well-defined away from the discontinuity at $x \simeq 0.0177$ while the values of $N_v$ fluctuate close to it. Assuming that a qualitatively similar behaviour will also occur for symmetric and Hermitean matrices, we now simplify the equations ${\cal P}=0$ associated with them. Retaining only five significant digits of the coefficients in these equations, we determine the number of MU vectors $\ket{\tilde v}$ and their inner products.    

Fig. \ref{Non-affine fig} shows that the family of \emph{symmetric} Hadamard matrices $M(t)$  comes with 48 MU vectors $\ket{\tilde v}$ close to the point $t=0$, while there are $120$ near $t=1/4$. These numbers are consistent with the rigorous bounds obtained in Sec. \ref{affinefamilies} if we recall that $M(0)=M(1/2)\approx F(0,0)$ and $M(1/4)\approx D(0)$ holds (cf. Fig. \ref{AllHadamards}). Across the entire parameter range, the number of MU vectors is a piecewise constant function symmetric about $x=1/4$, with distinct plateaus of $48, 52, 120$ and possibly 96 MU vectors. We suspect that the other values of $N_v$ near the discontinuities are spurious. An analysis of the scalar products among the approximate MU vectors shows that they can be arranged into bewteen $1$ and $16$ additional bases; a plot of which also resembles a step function. Crucially, they can never be arranged to form two bases that are MU to each other and therefore the points in Fig. \ref{Non-affine fig} cannot be included in a set of four MU bases. Table \ref{res:nonaffine} lists the results obtained for both a regular grid
\begin{equation} \label{Symmetricgrid}
\Gamma_{M} =\{a/144:a=1,2, \ldots , 71; a \neq 36\} \, ;
\end{equation}
and 300 randomly selected points in the fundamental interval; the reason for leaving out $a=36$ is the equivalence $M(1/4) \approx D_0$ just mentioned. We are confident that a more rigorous approach will confirm the absence of a set of four MU bases containing a single symmetric Hadamard matrix $M(t)$.  

The results obtained for \emph{Hermitean} Hadamard matrices $B(\theta)$, shown in Fig. \ref{Non-affine fig}, are similar to those of the symmetric family. The observed plateaus conform with the rigorous bounds found for $N_v=120$ and $N_v=56$  due to the equivalences $B(1/2) \approx D(0)$ and $B(\theta_0) \approx C$ (cf. Table \ref{res:special}). We consider the plateaus at 56, 58, 60, 72, 84 and 108 to be genuine while spurious values for $N_v$ proliferate near their ends, where $N_v$ is likely to vary discontinuously. Once more, Table \ref{res:nonaffine} reveals that both regularly spaced points on the grid
\begin{equation} \label{Hermiteangrid}
\Gamma_{B} =\{a/144:a=55,56, \ldots , 89; a \neq 72\} \, ;
\end{equation}
and randomly chosen values of the parameter $\theta$ define Hadamard matrices $B(\theta)$ which allow the construction of three MU bases but not four. 

Finally, let us consider the \emph{Sz\"oll\H{o}si} family, the non-affine two-parameter set of Ha\-da\-mard matrices  $X(a,b)$. Fig. \ref{szollosiCrossSection} shows the values of $N_v$ for randomly chosen parameters on two cuts through parameter space, namely along the line
\begin{equation} \label{lambdaline}
\Lambda =\{(a,b): \arg(a+ib)=\pi/6\} \, 
\end{equation}
which connects $X(0,0) \approx F(1/6,0)$ to the circulant matrix $C$, and the randomly chosen line
\begin{equation} \label{lambdaprimeline}
\Lambda ' =\{(a,b): \arg(a+ib)=0.3510\} \, 
\end{equation}
connecting $X(0,0)$ to $B(\theta^\prime)$, a Hermitean Hadamard matrix on the boundary. The values of $N_v$ at the end points of the lines are, in both cases, consistent with results obtained above for $F(1/6,0)$, $C$, and $B(\theta^\prime)$; broadly speaking, the number of solutions again represents a step function. However, the plateaus at $48, 52, 54, 56, 58$ and $60$ in Fig. \ref{szollosiCrossSection} (b) show considerable overlap: the effect of approximating the coefficients in the relevant polynomials is even more pronounced for the Sz\"oll\H{o}si family than for the other non-affine families. The results for the 300 randomly chosen parameter values sampling the two-dimensional parameter space resemble those of the symmetric and Hermitean families: we find $48 \leq N_v \leq 120$ throughout which allow for triples of MU bases but never for a quadruple. Preliminary calculations show that the properties of the new family of transposed Sz\"oll\H{o}si matrices $X^T(a,b)$ are similar to those of the set $X(a,b)$. 

While not being exact, the results for the symmetric, Hermitean and Sz\"oll\H{o}si families provide bounds on the number of MU bases which can be constructed from their members. None of the Hadamard matrices considered can be extended to a set of four MU bases. We consider it unlikely that the approximation made would systematically suppress other MU vectors with properties invalidating this conclusion.

\section{Summary and Conclusions}

We have searched for MU bases related to pairs $\{I,H\}$ where $I$ is the unit matrix and $H$ runs through a discrete subset of known ($6\times 6$) complex Hadamard matrices. Using Buchberger's algorithm, we have obtained upper bounds on the number of MU bases; the bounds are \emph{rigorous} in many cases and \emph{approximate} in others. Each of the 5,980 calculations required between 4 and 16 GB of memory and, altogether, would have lasted approximately 29,000 hours on a single 2.2 GHz processor.

Each point in Fig. \ref{AllHadamardsSampled} represents one of the Hadamard matrices $H$ we have been investigating. We find that the Spectral matrix $S$ is the only Hadamard matrix which cannot be extended to a triple of MU bases. Furthermore, if four (seven) MU bases were to exist in dimension six  three (six) Hadamard matrices different from the ones shown in Fig. \ref{AllHadamardsSampled} would be required. This clearly conforms with the numerically obtained evidence that no four MU basis exist \cite{brierley+08}. 

There is one caveat that we must make regarding the results for the non-affine families. In general, the program was unable to construct the associated Gr\"obner bases for the symmetric, Hermitean and Sz\"oll\H{o}si families. For these Hadamard matrices, we cannot guarantee that we have found \emph{all} MU vectors although we consider it unlikely that the approximation made would systematically suppress the missing vectors.

The symmetrical presentation of known Hadamard matrices in Fig \ref{AllHadamards} is justified by the results of our calculations: both the number of vectors $N_v$ and the values of their inner products (i.e. the number $N_t$) are symmetric about the line passing through $F(0,0)$ and $S$.  We will now explain why this symmetry exists. 

First, let $H$ be a member of the Di\c{t}\u{a}, symmetric or Fourier families and consider a vector $|v\rangle$ that is MU to both $I$ and $H$. Since multiplication by an overall unitary leaves the MU conditions (\ref{MUB conditions}) invariant, we have the equivalence between sets 
\begin{equation} \label{symmetry}
\{I,H,|v\rangle \}\approx \{H^{\dagger  },I,H^{\dagger  }|v\rangle \}
           = \{I,H^{\dagger  },|v^{\prime} \rangle\}
\end{equation}
where $|v^{\prime }\rangle =H^{\dagger  }|v\rangle$. It follows that $|v^{\prime }\rangle $ is MU to $I$ and $H^{\dagger  }$, and since $D^{\dagger  }(x)\approx D(-x)$,  $M^{\dagger  }(t)\approx M(-t)$ and $F^{\dagger 
}(x_{1},x_{2})\approx F^{T}(x_{1},x_{2})$, the number of solutions $N_v$ is
symmetric about the line through $F(0,0)$ and $S$. Further, since $H^{\dagger  }$ is
unitary and it is applied to all vectors, this transformation leaves the inner products between two MU vectors invariant and therefore, the number of triples $N_t$ is also symmetric.

We need an additional transformation to explain the symmetry found for the Hermitean matrices since $B^{\dagger  }(\theta)= B(\theta)$: under complex conjugation a Hermitean matrix transforms according to 
\begin{equation}
B^*(\theta) = B(1 -\theta ) \, ,
\end{equation}
as follows from the explicit form of $B(\theta)$ given in Eq. (\ref{defHermiteanFamily}). 
Now consider a vector $|v\rangle $ which is MU to the columns $b(\theta)$ of the matrix $B(\theta)$; then    
\begin{equation}
\left\vert \left\langle b(\theta)|v\right\rangle \right\vert ^{2}=\left\vert 
\left\langle b(\theta)|v\right\rangle^*\right\vert ^{2}=\left\vert
\left\langle b^*(\theta)| v^*\right\rangle \right\vert
^{2}=\left\vert \left\langle b(1-\theta)| v^* \right\rangle \right\vert
^{2}\, ,
\end{equation}
and therefore $| v^* \rangle $ is MU to each column of $B(1 -\theta )$. Thus, the vectors MU to $B(\theta )$ are the complex conjugates of those MU to $B(1 -\theta )$ which implies that the number $N_v$ of MU vectors (and their properties) will not change upon a reflection about the point $\theta=1/2$. Although we did not pay attention to the existence of these exact symmetries when introducing the approximations for the non-affine Hadamard matrices, the results obtained do respect them. 

The set of Hadamard matrices in $\mathbb{C}^6$ may depend on four parameters  \cite{bengtsson+07}, a conjecture which recently gained some numerical support \cite{skinner+08}. It remains difficult to draw general conclusions about the number of MU bases in dimension $d=6$. However, we would like to point out that the approach presented here is \emph{future-proof}: it will work for any Hadamard matrix---including currently unknown ones.  

In summary, we have shown that the construction of more than three MU bases in $\mathbb{C}^6$ is not possible starting from nearly 6,000 different Hadamard matrices. This result adds significant weight to the conjecture that a complete set of seven MU bases does not exist in dimension six. It becomes ever more likely that only prime-power dimensions allow for optimal state reconstruction.

\section*{Acknowledgments}

The calculations have been performed on the White Rose Grid provided by the
Universities of Leeds, Sheffield and York; we thank A. Turner and M. Hewitt, who run its node at York, for their help in using the grid.

\clearpage

\section*{Figures}

\begin{figure}[ht]
\begin{center}
\includegraphics[width=12cm]{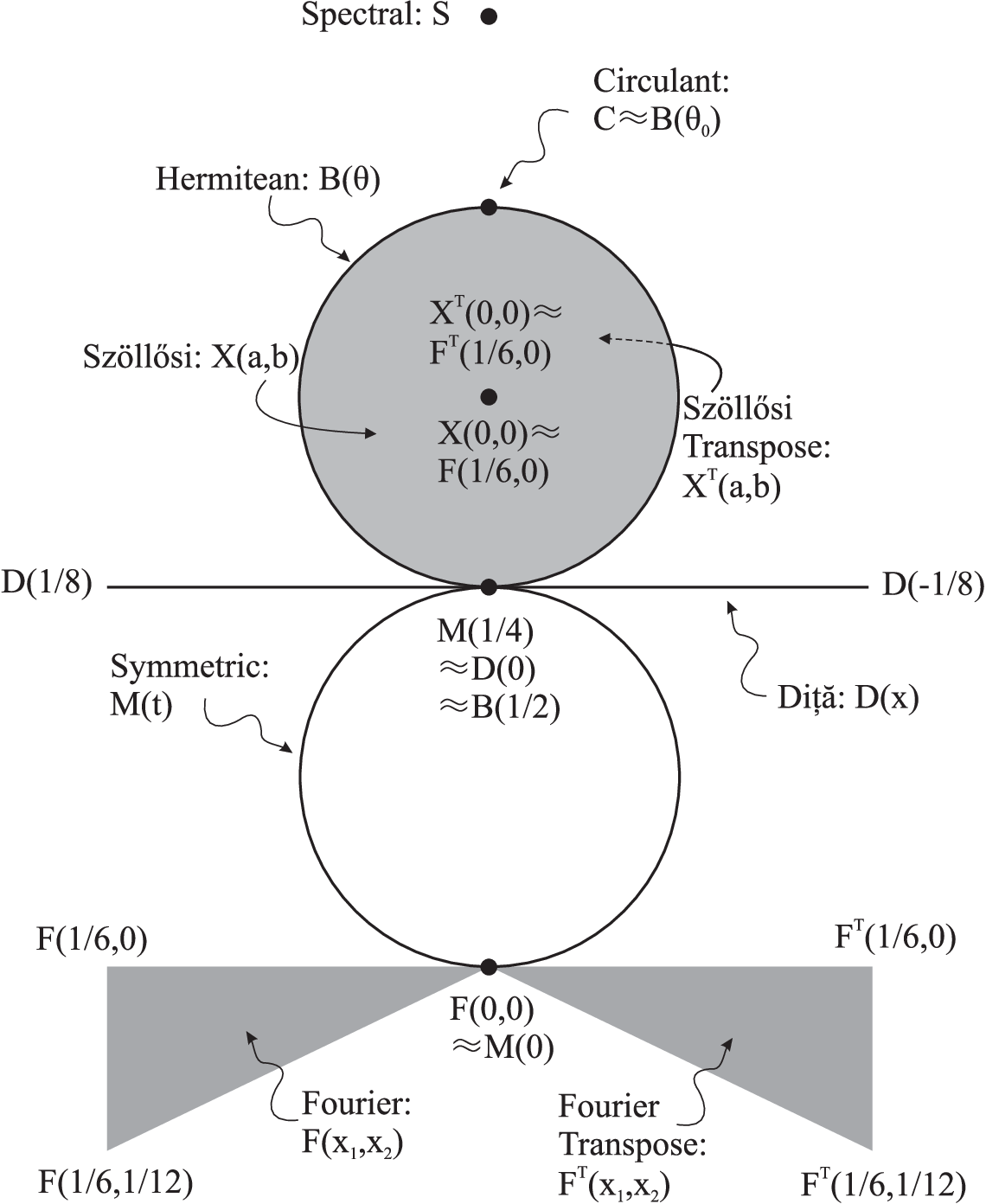}
\end{center}
\caption{The set of known Hadamard matrices in dimension six consists of  \emph{special} Hadamard matrices $F(0,0) \equiv F_6, D(0) \equiv D_0, C$, and $S$, located on the vertical symmetry axis; of the \emph{affine} families $D(x)$, $F({\bf x})$, and  $F^T({\bf x})$; and of the \emph{non-affine} families $M(t)$, $B(\theta)$, $X(a,b)$, and $X^T(a,b)$ (see Appendix \ref{defnHadamards} for definitions). Note that the sets $X(a,b)$ and $X^T(a,b)$ cover the interior of the upper circle twice.}
\label{AllHadamards}
\end{figure}

\begin{figure}[ht]
\begin{center}
\includegraphics[width=14cm]{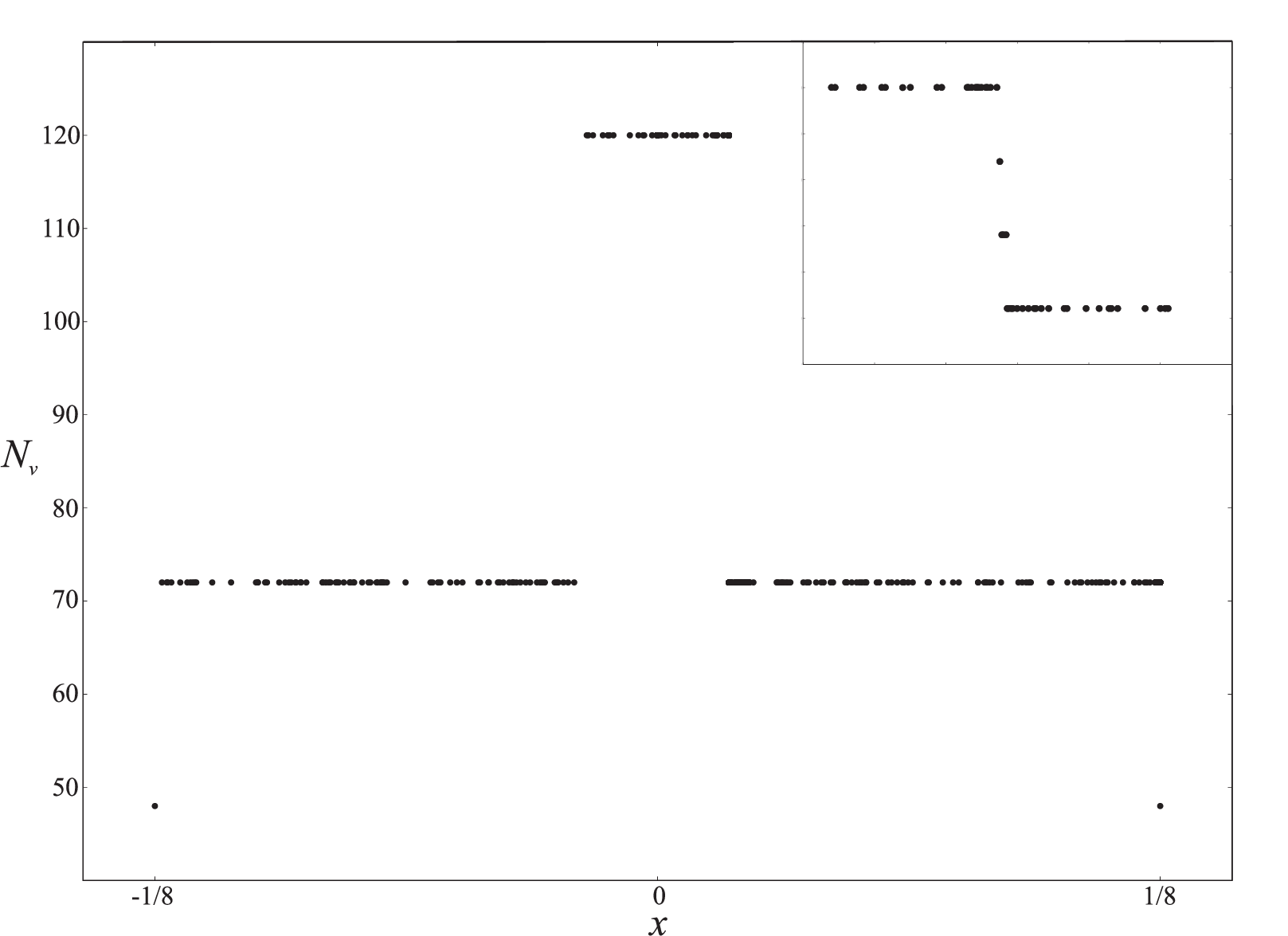}
\end{center}
\caption{The number $N_v$ of vectors $\ket{v}$ which are MU with respect to the columns of the identity $I$ and Di\c{t}\u{a} matrices $D(x)$; the parameter $x$ assumes 72 parameter values $x\in \Gamma_D$, and 500 randomly chosen ones in the fundamental interval $[-1/8,1/8]$ of the parameter $x$. The inset illustrates the impact on $N_v$ near the discontinuity $x \simeq 0.0177$ if an \emph{approximate} set of equations is used (cf. Sec. \ref{nonaffinefamilies}).}
\label{Dita fig}
\end{figure}

\begin{figure}[ht]
\begin{center}
\includegraphics[width=12.5cm]{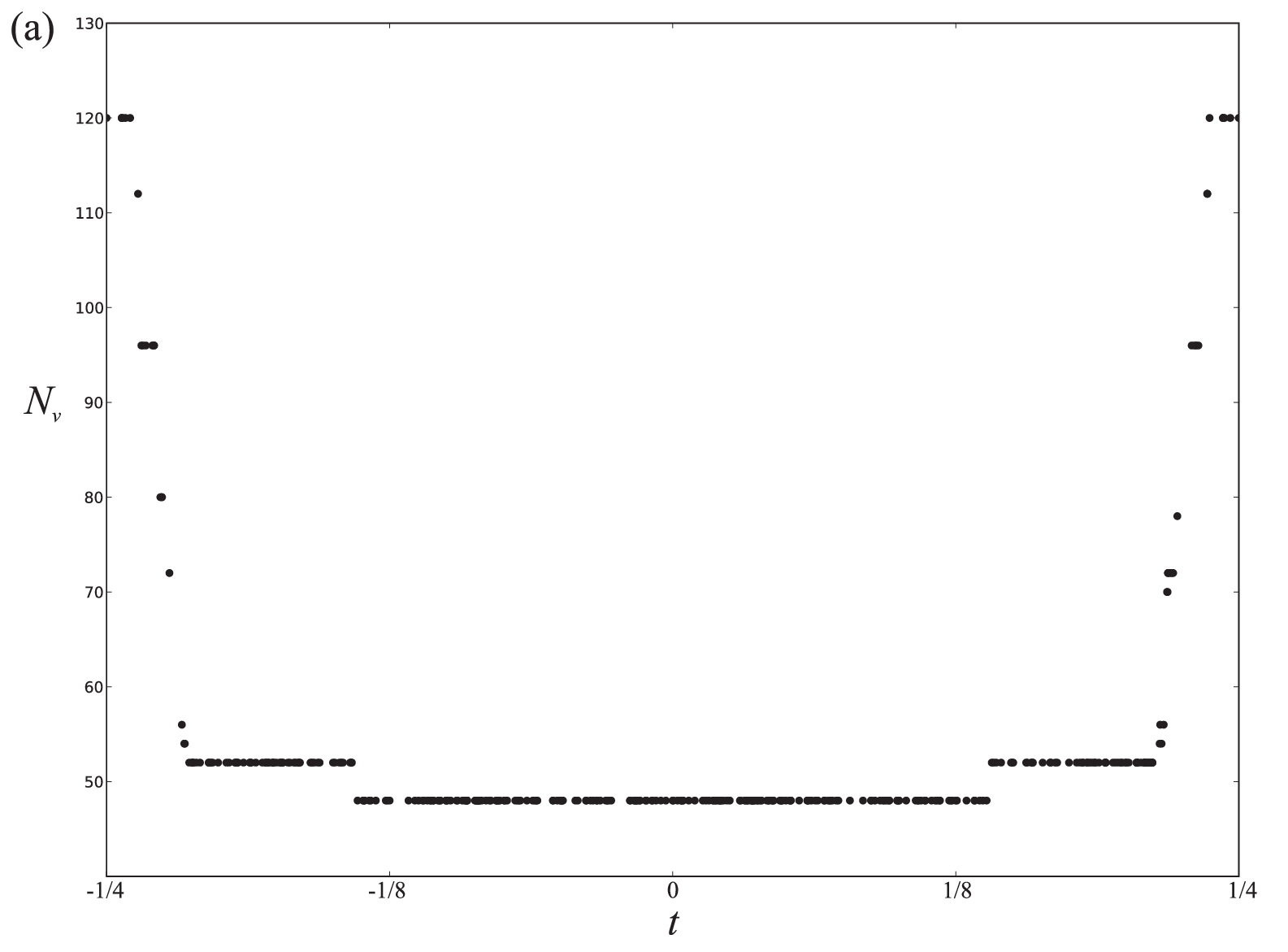}
\includegraphics[width=12.5cm]{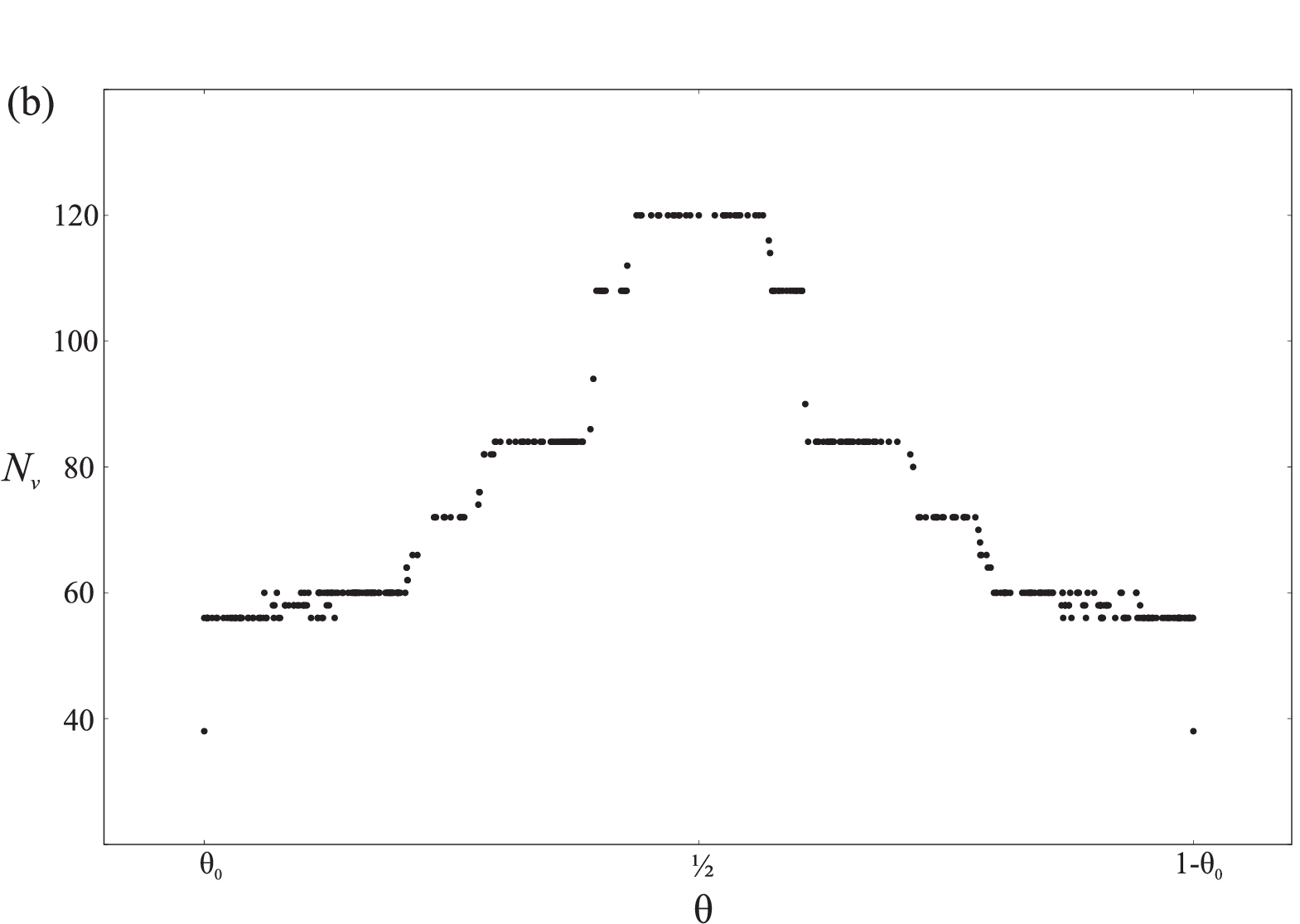}
\end{center}
\caption{The number $N_v$ of vectors $\ket{v}$ which are MU with respect to the columns of the identity $I$ and (a) symmetric Hadamard matrices $M(t)$; the parameter $t$ assumes 60 parameter values $t\in \Gamma_M$, and 300 randomly chosen ones in the fundamental interval $[0,1/2]$, and of (b) Hermitean matrices $B(\theta)$; the parameter $\theta$ assumes 34 parameter values $\theta\in \Gamma_B$, and 300 randomly chosen ones in the fundamental interval $[\theta_0,1-\theta_0]$. The phase $\theta_0$ has been defined in equation (\ref{theta0}).}
\label{Non-affine fig}
\end{figure}

\begin{figure}[ht]
\begin{center}
\includegraphics[width=13cm]{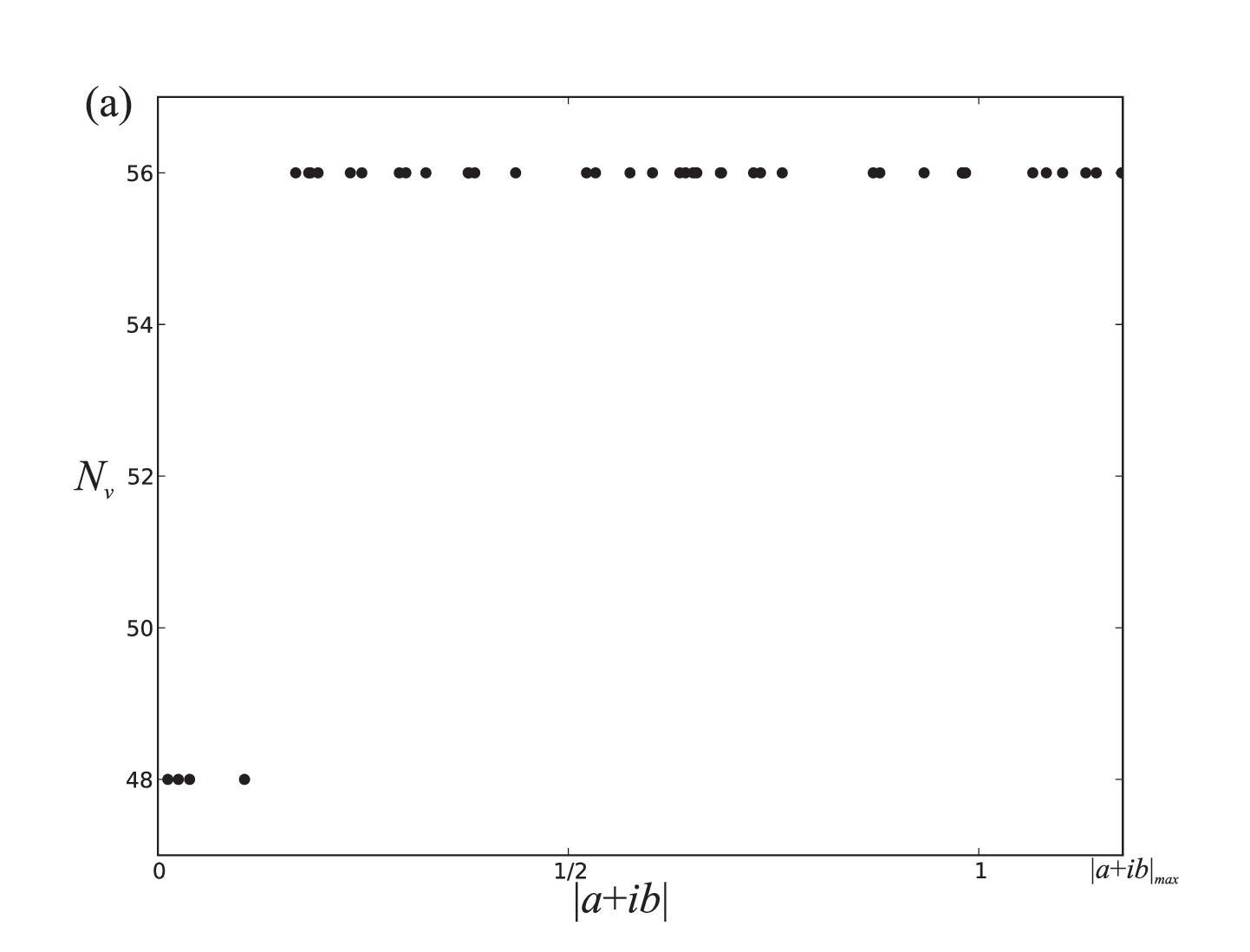}
\includegraphics[width=13cm]{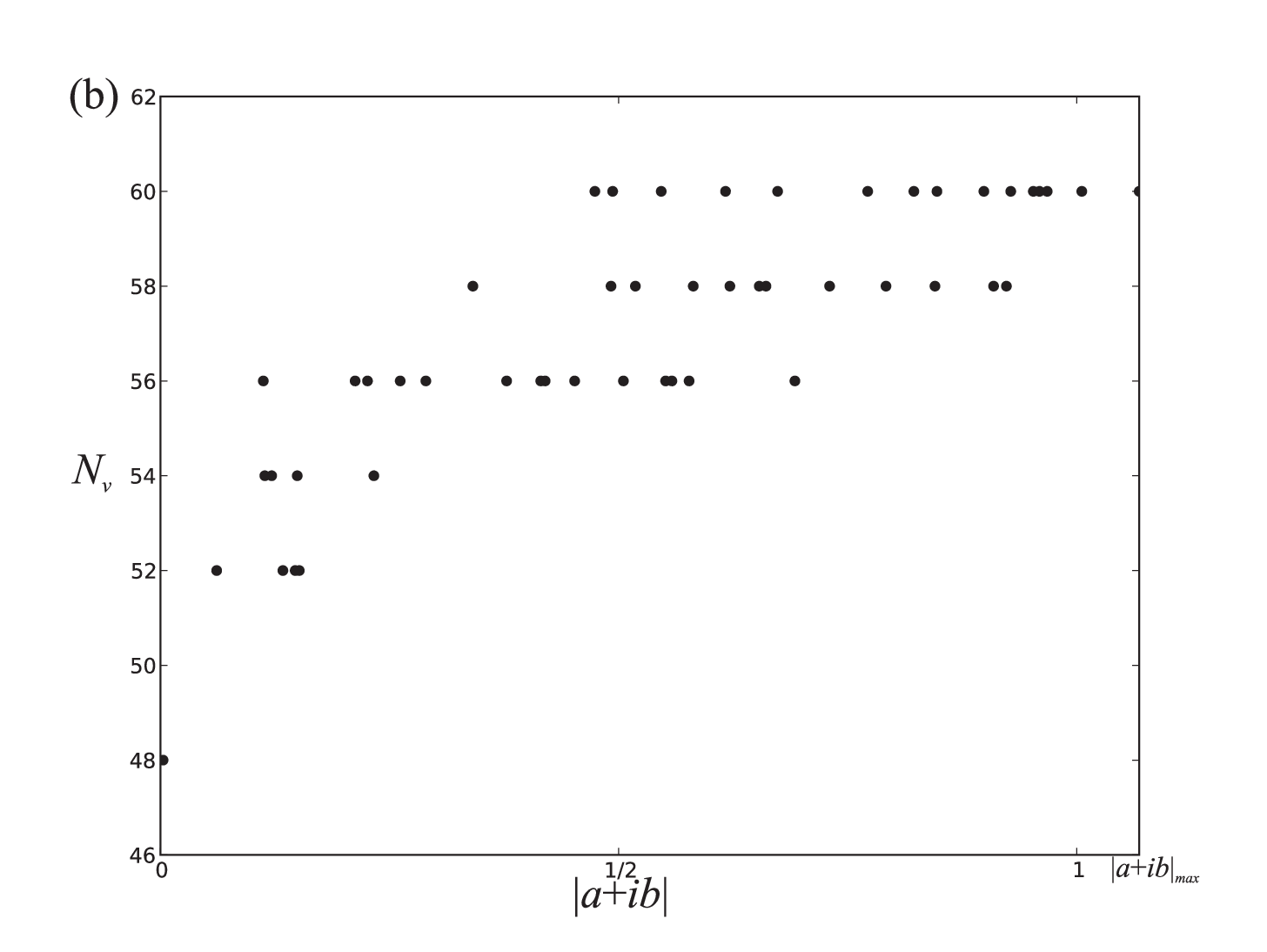}
\end{center}
\caption{The number $N_v$ of vectors $\ket{v}$ which are MU with respect to the columns of the identity $I$ and Sz\"{o}ll\H{o}si Hadamard matrices $X(a,b)$ for 50 randomly chosen parameter values (a) on the line $\Lambda$ connecting $F(1/6,0)$ to $C$, and (b) on the line $\Lambda^\prime$ connecting $F(1/6,0)$ to $B(\theta^\prime)$; in both figures, the maximum modulus $|a+ib|_{\footnotesize{\mbox{max}}}$ is defined by Eq. (\ref{deltoid}).}
\label{szollosiCrossSection}
\end{figure}

\begin{figure}[ht]
\begin{center}
\includegraphics[width=12cm]{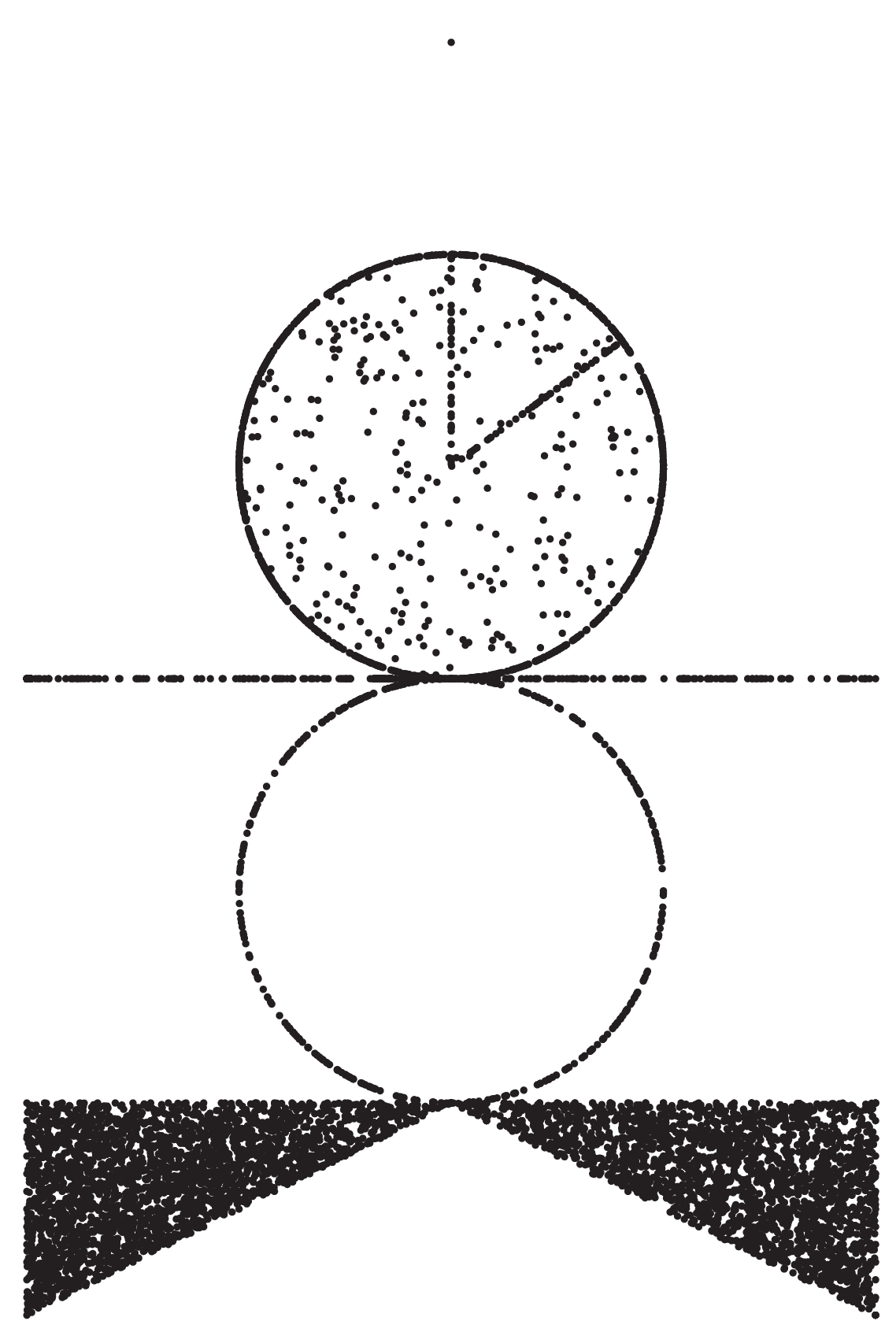}
\end{center}
\caption{The set of all Hadamard matrices $H$ which have been considered (cf. Fig. \ref{AllHadamards} and Tables \ref{res:special}-\ref{res:nonaffine}): for each $H$, a second MU Hadamard matrix can always be found except for the isolated spectral matrix $S$; consequently, triples of MU bases are the norm while quartets of MU bases do not exist.}
\label{AllHadamardsSampled}
\end{figure}

\clearpage

\section*{Tables}

\begin{table}[ht]
\begin{center}
\begin{tabular}{c|rr}
$H$  & $N_{v}$ & $N_{t}$  \\ \hline \hline
$F_6$ & 48 & 16 \\ 
$D_0$ & 120 & 10  \\
$C$ & 56 & 4  \\
$S$ & 90 & 0 
\end{tabular}
\end{center}
\caption{The number of MU vectors and their properties for \emph{special} Hadamard matrices: there are $N_{v}$ vectors being MU with respect to the pair of matrices $\{ I,H \}$ that form $N_{t}$ additional Hadamard matrices i.e. there are $N_t$ different \emph{triples} of MU bases.}
\label{res:special}
\end{table}

\begin{table}[h]
\begin{center}
\begin{tabular}{c|cr|rr}
$H$              & ${\bf x}$            & \#( ${\bf x}$) & $N_{v}$ & $N_{t}$  
\\ \hline \hline
$D(x)$        & $\Gamma_{D}$ & 36 & 48/72/120 & 4 \\ 
                 & random & 500 & 72/120 & 4  \\ \hline 
$F({\bf x})$     & $\Gamma_{F}$    & 168 & 48 & 8/70  \\ 
                 &random & 2,000 & 48 & 8  \\ \hline 
$F^T ({\bf x})$  & $\Gamma_{F}$    & 168 & 48 & 8/70  \\ 
                 & random & 2,000 & 48 & 8  \\ \hline
\end{tabular}%
\end{center}
\caption{The number of MU vectors and their properties for \emph{affine} Hadamard matrices: the second column indicates which values have been chosen for the parameters ${\bf x}$; the grids of points $\Gamma_{M}$ and $\Gamma_{F}$ are defined in Eqs. (\ref{Ditagrid}) and (\ref{Fouriergrid}), respectively; the third column displays the number of Hadamard matrices considered in a sample; $N_{v}$ and $N_{t}$ are defined as in Table \ref{res:special} and vary as a function of the parameter values (cf. Sec. \ref{affinefamilies}).} 
\label{res:affine}
\end{table}
\begin{table}[h!]
\begin{center}
\begin{tabular}{c|cr|rr}
$H$  & ${\bf x}$        & \#( ${\bf x}$) & $N_{v}$ & $N_{t}$  
 \\ \hline \hline
$M(t)$ & $\Gamma_{M}$     & 70            & 48-120  & 1-16          \\ 
     & random           & 300           & 48-120  & 1-16        \\ \hline
 $B(\theta)$ &  $\Gamma_{B}$ & 34       & 56-120  & 1/4/8/16         \\ 
             & random   & 300           & 56-120  & 1/4/8/16         \\
\hline
$X(a,b)$	& $\Lambda$ & 50 & 48/56 & 4/16/70  \\
			& $\Lambda^\prime$ & 50 & 48-60 & 4/8/16/70  \\
			& random & 300 & 48-120 & 1-70     \\
\end{tabular}%
\end{center}
\caption{The number of MU vectors and their properties for \emph{non-affine} Hadamard matrices: the grids $\Gamma_{M}$ and $\Gamma_{B}$ are defined in Eqs. (\ref{Symmetricgrid}) and (\ref{Hermiteangrid}), respectively; see Eqs. (\ref{lambdaline}) and (\ref{lambdaprimeline}) for the definition of the lines $\Lambda$ and $\Lambda^\prime$; other notation as in Table \ref{res:affine}; preliminary results for the family $X^T(a,b)$ resemble those obtained for $X(a,b)$.}
\label{res:nonaffine}
\end{table}

\newpage \appendix

\section{Known complex Hadamards matrices in dimension six} 
\label{defnHadamards}

This Appendix lists the currently known complex Hadamard matrices for easy reference and to establish notation. For more details the reader is referred to \cite{bengtsson+07} and to the online catalogue \cite{Hadamardsonline}.  

\subsection{Special Hadamard matrices} \label{specialApp}

The \emph{Fourier matrix} $F_6$ has been introduced in Eq. (\ref{F6}); it is contained in both the Fourier family $F({\bf x})$ and the transposed Fourier family $F^T({\bf x})$ for ${\bf x} = 0$, where $F_6 \equiv F(0,0)\approx F^{T}(0,0)$ holds (cf. Sec. \ref{affineApp}).

The \emph{Di\c{t}\u{a} matrix} $D_0$ is an example of a complex symmetric Hadamard matrix, 
\begin{equation} \label{defDita0}
D_{0}=\frac{1}{\sqrt{6}}\left( 
\begin{array}{cccccc}
1 & 1 & 1 & 1 & 1 & 1 \\ 
1 & -1 & i & -i & -i & i \\ 
1 & i & -1 & i & -i & -i \\ 
1 & -i & i & -1 & i & -i \\ 
1 & -i & -i & i & -1 & i \\ 
1 & i & -i & -i & i & -1%
\end{array}%
\right) \, ,
\end{equation}
embedded in a continuous one-parameter set of Hadamard matrices, the Di\c{t}\u{a} family (cf. \ref{affineApp}).

Bj\"orck's \emph{circulant matrix} \cite{bjorck+95} is defined by
\begin{equation}
C=\frac{1}{\sqrt{6}}\left( 
\begin{array}{cccccc}
1 & iz & -z & -i & -z^* & iz^* \\ 
i z^* & 1 & iz & -z & -i & -z^* \\ 
-z^* & iz^* & 1 & iz & -z & -i \\ 
-i & -z^* & iz^* & 1 & iz & -z \\ 
-z & -i & -z^* & iz^* & 1 & iz \\ 
iz & -z & -i & -z^* & iz^* & 1%
\end{array}
\right) \, ,
\end{equation}
where 
\begin{equation}
z=\frac{1-\sqrt{3}}{2}+i\sqrt{\frac{\sqrt{3}}{2}} \, .
\end{equation}
It was originally thought to be isolated but it is now known to be part of the family of Hermitean Hadamard matrices,  $C\approx B(\theta _{0})$ (cf. \ref{nonaffApp}).
 
The only known isolated Hadamard matrix is the \emph{spectral matrix},  
\begin{equation}
S=\left( 
\begin{array}{cccccc}
1 & 1 & 1 & 1 & 1 & 1 \\ 
1 & 1 & \omega & \omega & \omega ^{2} & \omega ^{2} \\ 
1 & \omega & 1 & \omega ^{2} & \omega ^{2} & \omega \\ 
1 & \omega & \omega ^{2} & 1 & \omega & \omega ^{2} \\ 
1 & \omega ^{2} & \omega ^{2} & \omega & 1 & \omega \\ 
1 & \omega ^{2} & \omega & \omega ^{2} & \omega & 1%
\end{array}
\right) \, ,
\end{equation}
where $\omega $ is a third root of unity, $\omega = e^{2\pi i/3}$. It has been discovered by Moorhouse \cite{moorhouse01} and, independently, by Tao \cite{tao04}.

\subsection{Affine Families} \label{affineApp}

There are three affine families of Hadamard matrices, characterized by the property (\ref{defAffine}) that they can be written as a non-trivial Hadamard product. The  \emph{Di\c{t}\u{a} family} \cite{dita04} is given by $D(x)=D_{0} \circ \text{Exp}[2\pi iR(x)]$, $|x| \leq 1/8$, with $D_0$ from Eq. (\ref{defDita0}) and 
\begin{equation}
 R(x)=\left( 
\begin{array}{cccccc}
0 & 0 & 0 & 0 & 0 & 0 \\ 
0 & 0 & 0 & 0 & 0 & 0 \\ 
0 & 0 & 0 & x & x & 0 \\ 
0 & 0 & -x & 0 & 0 & -x \\ 
0 & 0 & -x & 0 & 0 & -x \\ 
0 & 0 & 0 & x & x & 0%
\end{array}
\right) \, ;
\end{equation}
the componentwise exponential $\mbox{Exp}[\cdot] $ of a matrix has been defined after Eq. (\ref{defAffine}). 

The Fourier matrix $F_6$ has been embedded in a similar way into a \emph{two}-parameter set, namely the Fourier family $F({\bf x})=F_6 \circ \text{Exp}[2\pi iR({\bf x})]$, where  
\begin{equation}
R({\bf x}) \equiv R(x_{1},x_{2})=\left( 
\begin{array}{cccccc}
0 & 0 & 0 & 0 & 0 & 0 \\ 
0 & x_{1} & x_{2} & 0 & x_{1} & x_{2} \\ 
0 & 0 & 0 & 0 & 0 & 0 \\ 
0 & x_{1} & x_{2} & 0 & x_{1} & x_{2} \\ 
0 & 0 & 0 & 0 & 0 & 0 \\ 
0 & x_{1} & x_{2} & 0 & x_{1} & x_{2}%
\end{array}
\right)\, ;
\end{equation}
the parameters $(x_1, x_2)$ take values in a fundamental region given by a triangle with vertices $(0,0),$ $(1/6,0)$ and $(1/6,1/12)$.

Upon transposing the matrices $F({\bf x})$ one obtains a different two-parameter set of Hadamard matrices, called the \emph{transposed Fourier family} $F^T({\bf x})$. It has the same fundamental region as the Fourier family.

\subsection{Non-Affine Families} \label{nonaffApp}

Non-affine Hadamard matrices are not parametrised in the form (\ref{defAffine}).   The \emph{Hermitean family} \cite{beauchamp+06} provides a one-parameter example of such a set,
\begin{equation} \label{defHermiteanFamily}
B(\theta )=\frac{1}{\sqrt{6}}\left( 
\begin{array}{cccccc} 
1 & 1 & 1 & 1 & 1 & 1 \\ 
1 & -1 & x^* & -y & y & x^* \\ 
1 & -x & 1 & y & z^* & t^* \\ 
1 & y^* & y^* & 1 & t^* & t^* \\ 
1 & y^* & z & -t & 1 & x^* \\ 
1 & x & -t & t & -x & 1%
\end{array}%
\right) \, ,
\end{equation}%
where $y=e^{2\pi i\theta }$ and $t=xyz,$ with 
\begin{eqnarray*}
z &=&\frac{1+2y-y^{2}}{y(-1+2y+y^{2})}\, , \\
x &=&\frac{1+2y+y^{2}\pm \sqrt{2(1+2y+2y^{3}+y^{4})}}{1+2y-y^{2}} \, ;
\end{eqnarray*}%
the free parameter $\theta$ is restricted to vary within the fundamental interval   $ \lbrack \theta _{0},1-\theta _{0}]$, and the number $\theta_0$ is defined by the condition
\begin{equation} 
2\pi \theta _{0}=\cos ^{-1}\left( 1-\sqrt{3} \right) \, .
\label{theta0}
\end{equation}
Note that this is a smaller fundamental region than was previously known; the reduction is due to equivalences that have become apparent since the discovery of the Sz\"{o}ll\H{o}si family (cf. below).

Another non-affine one-parameter set of Hadamard matrices is given by the \emph{symmetric family} \cite{matolcsi+07},
\begin{equation}
M(t)=\frac{1}{\sqrt{6}}\left( 
\begin{array}{cccccc}
1 & 1 & 1 & 1 & 1 & 1 \\ 
1 & -1 & x & x & -x & -x \\ 
1 & x & d & a & b & c \\ 
1 & x & a & d & c & b \\ 
1 & -x & b & c & p & q \\ 
1 & -x & c & b & q & p%
\end{array}%
\right)\,  ,
\end{equation}%
where $x=e^{2\pi it}$, and the complex numbers $a,b,c,d,p,q$ are the unique solutions of the equations 
\begin{eqnarray}
1+x+d+a+b+c &=&0 \, , \nonumber\\
x^{2}-2x-2a-2d-1 &=&0 \, , \nonumber\\
1-x+b+c+p+q &=&0 \, , \nonumber\\
x^{2}+2b+2c+1 &=&0 \, . \label{symmmetricfamilyconditions}
\end{eqnarray}
In addition, one needs the fact that given a row $(r_{1},\ldots ,r_{6})$ of a Hadamard matrix, the last two elements are determined by $\Sigma = (r_{1}+r_{2}+r_{3}+r_{4})/2$, since 
\begin{equation}
r_{5,6}=-\Sigma \pm i\frac{\Sigma }{|\Sigma |}\sqrt{1-|\Sigma |^{2}}
\end{equation}%
if $\Sigma \neq 0.$ The fundamental region is given by $t\in \lbrack 0,1/2]$.

Finally, there is the non-affine \emph{Sz\"{o}ll\H{o}si family} \cite{szollosi08}
\begin{equation}
X(a,b)\equiv H(x,y,u,v)=\frac{1}{\sqrt{6}}\left( 
\begin{array}{cccccc}
1 & 1 & 1 & 1 & 1 & 1 \\ 
1 & x^2y & xy^2 & \frac{xy}{uv} & uxy & vxy \\ 
1 & \frac{x}{y} & x^2y & \frac{x}{u} & \frac{x}{v} & uvx \\ 
1 & uvx & uxy & -1 & -uxy & -uvx \\ 
1 & \frac{x}{u} & vxy & -\frac{x}{u} & -1 & -vxy \\ 
1 & \frac{x}{v} & \frac{xy}{uv} & -\frac{xy}{uv} &-\frac{x}{v} &  -1%
\end{array}%
\right)\, .
\end{equation}%
The entries $x$, $y$ and $u$, $v$ are solutions to the equations $f_{\alpha}=0$ and $f_{-\alpha}=0$, respectively, where
\begin{equation}\label{SzollosiEq}
f_{\alpha}(z) \equiv z^3- \alpha z^2+\alpha^{*} z-1 \, ,
\end{equation}
and $\alpha \equiv a+ib$ is restricted to the region $\mathbb{D}$ defined by $D(\alpha)\leq 0$ and $D(-\alpha)\leq 0$, with
\begin{equation}\label{deltoid}
D(\alpha)\equiv|\alpha|^4+18|\alpha|^2-8 \mbox{Re}[\alpha^3]-27 \, .
\end{equation}
It is possible to reduce $\mathbb{D}$ to a smaller fundamental region \cite{bengtsson08} since, firstly, the transformation $\alpha\rightarrow -\alpha$ maps Hadamard matrices to equivalent ones and, second, Eq. (\ref{SzollosiEq}) is invariant under the substitutions $\alpha \rightarrow \omega \alpha$ and $y \rightarrow \omega y$ with $\omega=\exp(2\pi i/3)$. As the second transformation leaves the dephased Hadamard matrix invariant, this establishes an equivalence between the Hadamard matrices associated with points in $\mathbb{D}$ and in $\mathbb{D}'$ (which one obtains from $\mathbb{D}$ through a rotation by $2 \pi/3$). As a result, the region $\mathbb{D}$ is found to consist of six equivalent sectors, and one may restrict $\alpha$ by 
\begin{equation}
0\leq  \arg(\alpha)\leq \frac{\pi}{3} \, .
\end{equation}

The \emph{transposed Sz\"{o}ll\H{o}si family}, $X^T(a,b)$ is obtained by transposing $X(a,b)$ or by using the equivalence $H(x,y,u,v)^T \approx H(x,y,v,u)$.
Fig. \ref{AllHadamards} illustrates that the points on the boundary of the reduced fundamental region for both $X(a,b)$ and $X^T(a,b)$ correspond to the members of the Hermitean family. 

\section{Simplification of the Fourier equations in dimension 6\label%
{SimplifyEqns}}

The conditions for a state $\ket{v} \in \mathbb{C}^6$  to be MU with respect to $F_6$ are given by ${\cal P}=0$ where  ${\cal P}=\{p_{\pm },q_{\pm},r_{\pm }\}$ with
\begin{eqnarray}
p_{\pm } &=&-5\pm 2\,x_{{5}}+2\,x_{{4}}\pm 2\,x_{{3}}+2\,x_{{2}}\pm 2\,x_{{1}%
}+{x_{{5}}}^{2}\pm 2\,x_{{4}}x_{{5}}+{x_{{4}}}^{2}+2\,x_{{3}}x_{{5}}\pm
2\,x_{{3}}x_{{4}} \nonumber \\
&&+{x_{{3}}}^{2}\pm 2\,x_{{2}}x_{{5}}+2\,x_{{2}}x_{{4}}\pm 2\,x_{{2}}x_{{3}}+%
{x_{{2}}}^{2}+2\,x_{{1}}x_{{5}}\pm 2\,x_{{1}}x_{{4}}+2\,x_{{1}}x_{{3}}\pm
2\,x_{{1}}x_{{2}} \nonumber\\
&&+{x_{{1}}}^{2}+{y_{{5}}}^{2}\pm 2\,y_{{4}}y_{{5}}+{y_{{4}}}^{2}+2\,y_{{3}%
}y_{{5}}\pm 2\,y_{{3}}y_{{4}}+{y_{{3}}}^{2}\pm 2\,y_{{2}}y_{{5}}+2\,y_{{2}%
}y_{{4}}\pm 2\,y_{{2}}y_{{3}} \nonumber \\
&&+{y_{{2}}}^{2}+2\,y_{{1}}y_{{5}}\pm 2\,y_{{1}}y_{{4}}+2\,y_{{1}}y_{{3}}\pm
2\,y_{{1}}y_{{2}}+{y_{{1}}}^{2}\, , \nonumber \\
q_{\pm } &=&-5+x_{{5}}-x_{{4}}-2\,x_{{3}}-x_{{2}}+x_{{1}}\mp \sqrt{3}y_{{5}%
}\mp \sqrt{3}y_{{4}}\pm \sqrt{3}y_{{2}}\pm \sqrt{3}y_{{1}}+{x_{{5}}}^{2}+x_{{%
4}}x_{{5}} \nonumber\\
&&+{x_{{4}}}^{2}-x_{{3}}x_{{5}}+x_{{3}}x_{{4}}+{x_{{3}}}^{2}-2\,x_{{2}}x_{{5}%
}-x_{{2}}x_{{4}}+x_{{2}}x_{{3}}+{x_{{2}}}^{2}-x_{{1}}x_{{5}}-2\,x_{{1}}x_{{4}%
} \nonumber\\
&&-x_{{1}}x_{{3}}+x_{{1}}x_{{2}}+{x_{{1}}}^{2}\pm \sqrt{3}y_{{5}}x_{{4}}\pm 
\sqrt{3}y_{{5}}x_{{3}}\mp \sqrt{3}y_{{5}}x_{{1}}+{y_{{5}}}^{2}\mp \sqrt{3}y_{%
{4}}x_{{5}}\pm \sqrt{3}y_{{4}}x_{{3}} \nonumber\\
&&\pm \sqrt{3}y_{{4}}x_{{2}}+y_{{4}}y_{{5}}+{y_{{4}}}^{2}\mp \sqrt{3}y_{{3}%
}x_{{5}}\mp \sqrt{3}y_{{3}}x_{{4}}\pm \sqrt{3}y_{{3}}x_{{2}}\pm \sqrt{3}y_{{3%
}}x_{{1}}-y_{{3}}y_{{5}} \nonumber\\
&&+y_{{3}}y_{{4}}+{y_{{3}}}^{2}\mp \sqrt{3}y_{{2}}x_{{4}}\mp \sqrt{3}y_{{2}%
}x_{{3}}\pm \sqrt{3}y_{{2}}x_{{1}}-2\,y_{{2}}y_{{5}}-y_{{2}}y_{{4}}+y_{{2}%
}y_{{3}}+{y_{{2}}}^{2} \nonumber\\
&&\pm \sqrt{3}y_{{1}}x_{{5}}\mp \sqrt{3}y_{{1}}x_{{3}}\mp \sqrt{3}y_{{1}}x_{{%
2}}-y_{{1}}y_{{5}}-2\,y_{{1}}y_{{4}}-y_{{1}}y_{{3}}+y_{{1}}y_{{2}}+{y_{{1}}}%
^{2}\, , \nonumber \\
r_{\pm } &=&-5-x_{{5}}-x_{{4}}+2\,x_{{3}}-x_{{2}}-x_{{1}}\mp \sqrt{3}y_{{5}%
}\mp \sqrt{3}y_{{4}}\pm \sqrt{3}y_{{2}}\mp \sqrt{3}y_{{1}}+{x_{{5}}}^{2}-x_{{%
4}}x_{{5}} \nonumber\\
&&+{x_{{4}}}^{2}-x_{{3}}x_{{5}}-x_{{3}}x_{{4}}+{x_{{3}}}^{2}+2\,x_{{2}}x_{{5}%
}-x_{{2}}x_{{4}}-x_{{2}}x_{{3}}+{x_{{2}}}^{2}-x_{{1}}x_{{5}}+2\,x_{{1}}x_{{4}%
} \nonumber \\
&&-x_{{1}}x_{{3}}-x_{{1}}x_{{2}}+{x_{{1}}}^{2}\pm \sqrt{3}y_{{5}}x_{{4}}\mp 
\sqrt{3}y_{{5}}x_{{3}}\pm \sqrt{3}y_{{5}}x_{{1}}+{y_{{5}}}^{2}\mp \sqrt{3}y_{%
{4}}x_{{5}}\pm \sqrt{3}y_{{4}}x_{{3}} \nonumber \\
&&\mp \sqrt{3}y_{{4}}x_{{2}}-y_{{4}}y_{{5}}+{y_{{4}}}^{2}\pm \sqrt{3}y_{{3}%
}x_{{5}}\mp \sqrt{3}y_{{3}}x_{{4}}\pm \sqrt{3}y_{{3}}x_{{2}}\mp \sqrt{3}y_{{3%
}}x_{{1}}-y_{{3}}y_{{5}} \nonumber \\
&&-y_{{3}}y_{{4}}+{y_{{3}}}^{2}\pm \sqrt{3}y_{{2}}x_{{4}}\mp \sqrt{3}y_{{2}%
}x_{{3}}\pm \sqrt{3}y_{{2}}x_{{1}}+2\,y_{{2}}y_{{5}}-y_{{2}}y_{{4}}-y_{{2}%
}y_{{3}}+{y_{{2}}}^{2} \nonumber \\
&&\mp \sqrt{3}y_{{1}}x_{{5}}\pm \sqrt{3}y_{{1}}x_{{3}}\mp \sqrt{3}y_{{1}}x_{{%
2}}-y_{{1}}y_{{5}}+2\,y_{{1}}y_{{4}}-y_{{1}}y_{{3}}-y_{{1}}y_{{2}}+{y_{{1}}}^{2}\, . \label{pqr}
\end{eqnarray}
Upon substituting the normalization condition $\bra{v}v\rangle =1$, or 
\begin{equation}
{x_{{1}}}^{2}+{y_{{1}}}^{2}+{x_{{2}}}^{2}+{y_{{2}}}^{2}+{x_{{3}}}^{2}+{y_{{%
3}}}^{2}+{x_{{4}}}^{2}+{y_{{4}}}^{2}+{x_{{5}}}^{2}+{y_{{5}}}^{2}=5 \, ,
\end{equation}
one finds 
\begin{eqnarray}
p_{+}+p_{-} &=& 0\, , \nonumber \\
p_{+}-\,p_{-}-q_{+}+q_{-}+r_{+}-r_{-} &=& 0\, , \nonumber\\
2p_{+}-2\,p_{-}+q_{+}-q_{-}-r_{+}+r_{-} &=& 0\, , \nonumber\\
p_{+}\pm \,p_{-}\mp r_{+}-r_{-} &=& 0\, , \label{simplify}
\end{eqnarray}
giving Eqs. (\ref{F00}). 

\end{document}